# Visualizing a Field of Research: A Methodology of Systematic Scientometric Reviews

*Chaomei Chen[1,2] and Min Song[2]*
*[1]Department of Information Science, College of Computing and Informatics, Drexel University*
*[2]Department of Information Science, Yonsei University*

Systematic scientometric reviews, empowered by scientometric and visual analytic techniques, offer opportunities to improve the timeliness, accessibility, and reproducibility of conventional systematic reviews. While increasingly accessible science mapping tools enable end users to visualize the structure and dynamics of a research field, a common bottleneck in the current practice is the construction of a collection of scholarly publications as the input of the subsequent scientometric analysis and visualization. End users often have to face a dilemma in the preparation process: the more they know about a knowledge domain, the easier it is for them to find the relevant data to meet their needs adequately; the little they know, the harder the problem is. What can we do to avoid missing something valuable but beyond our initial description? In this article, we introduce a flexible and generic methodology, cascading citation expansion, to increase the quality of constructing a bibliographic dataset for systematic reviews. Furthermore, the methodology simplifies the conceptualization of globalism and localism in science mapping and unifies them on a consistent and continuous spectrum. We demonstrate an application of the methodology to the research of literature-based discovery and compare five datasets constructed based on three use scenarios, namely a conventional keyword-based search (one dataset), an expansion process starting with a groundbreaking article of the knowledge domain (two datasets), and an expansion process starting with a recently published review article by a prominent expert in the domain (two datasets). The unique coverage of each of the datasets is inspected through network visualization overlays with reference to other datasets in a broad and integrated context.

## Introduction

Systematic reviews play a critical role in scholarly communication (1). Systematic reviews typically synthesize findings from original research in a field of study, assess the degree of consensus or the lack of it concerning the state of the art in the field, and identify challenges and future directions. For newcomers to a field of study, a timely and comprehensive systematic review can provide a valuable overview of the intellectual landscape such that new researchers may find research topics to pursue effectively. For experienced and active researchers, systematic reviews can be instrumental in keeping their knowledge of the field up to date, especially when involving areas that are potentially relevant but fall outside the immediate topic of one's interest. For other stakeholders of a scientific field, such as policy makers, administrators of research institutions, and the public, systematic reviews may help them to develop a good understanding of science in the context of their everyday life.

A fast-growing trend is the increase of systematic reviews conducted with the assistance of science mapping tools (2). A science mapping tool typically takes a set of bibliographic records of a research field and generates an overview of the underlying knowledge domain, e.g. as with CiteSpace (3, 4) and VOSviewer (5). For example, systematic reviews facilitated by using CiteSpace include research areas such as regenerative medicine (6), natural disaster research (7), greenhouse gas emission (8), and identifying disruptive innovation and emerging technology (9). Similarly, VOSviewer was used in reviews of topics such as citizen science (10) and climate change (11). Using an overview of the intellectual landscape of a field of study provides a practical remedy to situations in which a good systematic review is not readily available or accessible. Furthermore, given that domain expertise is always a valuable resource when it comes to assess the state of the art in a domain, computational tools start to shift the burden from human domain experts to computational techniques that can extract meaningful information from text documents and synthesize various thematic patterns. Scientometric studies can provide domain-specific insights without direct involvements of domain experts (6).

Researchers differentiate science mapping efforts in terms of global, local, and hybrid science maps (12). Global maps of science by definition provide a comprehensive coverage of all scientific disciplines (13), whereas local maps typically focus on some areas of research but are not expected to represent all areas. Hybrid maps may use a global map as a base map and superimpose local maps as overlays (14). Generating global maps of science requires a substantial array of resources that are not commonly accessible to the majority of the research community, whereas local maps are more commonly seen in published review articles. Global maps tend to focus on disciplines and interdisciplinary relations at a higher level of granularity than the article-level or topic-level granularity in local maps. In practice, generating global maps is largely limited to a small number of researchers with direct access to large databases of scientific publications. Consequently, global maps may not be generated and updated as often as individual researchers wish for and they may not be tailored to specific needs of individual researchers. In contrast, the most distinct advantage of generating local maps is the accessibility of tools; individual researchers are in control of the fast-growing pool of science mapping tools, notably CiteSpace (3, 4), VOSviewer (5), and HistCite (15). They can choose when and what subject area to work with as they wish.

Being able to choose a subject area to work with freely in a local map approach can be an advantage or a challenge. On the one hand, the user is now responsible for selecting the data as the input. On the other hand, we all want to learn something new and something that we don't even know at the beginning of the survey. How can we specify things we don't even know? This is a dilemma, especially for researchers who are new to a field of research. A commonly used strategy is to enlist domain experts to construct a search query as comprehensive as possible. This strategy may have a few drawbacks. The user may end up with a long and complex query. Furthermore, the query may be biased by the very expertise of the domain experts. More importantly, this strategy still does not resolve the dilemma. Another useful but potentially time-consuming strategy is to follow an iterative process, which is in principle similar to the idea of rapid prototyping in software engineering. The researcher would conduct multiple iterations of science mapping with the aim to learn from each iteration about the target field so as to improve the query for the next iteration. Ideally, the process will converge or become stable after several iterations. In practice, however, this strategy may entail a considerable amount of overhead and it is likely to be time consuming. In fact, this process is a miniature of conducting research in general. The expected convergence may never come. This dilemma underlines the most fundamental challenge to the adaptation of a local map approach because the quality of the input data has a direct impact on the overall quality of the systematic review to be built on the basis of the overview of a field of interest.

In this article, we propose a generic computational approach to science mapping that generates systematic reviews of a research field. The new approach automates the essence of the multi-iteration strategy. Starting with an initial dataset, the process will automatically expand the initial set by retrieving additional records of relevant publications. The expansion can move forward into the future and backward into the past with reference to a given publication. This flexibility could be particularly valuable in some of the most commonly encountered scenarios of research. For example, one may start the process with a groundbreaking article of a field and collect relevant and subsequently published articles. In contrast, one may start with a recently published systematic review of a field and collect relevant articles published earlier on. Pragmatically speaking, the ability to expand a given collection of scholarly publications provides a smooth transition between a local map approach and a global map approach. The transitional approach enhances the capability of science mapping tools in the hand of end-users and reduces the burden on the end user to formulate a comprehensive and representative collection of bibliographic records as the input of a systematic review.

We demonstrate how this expansion procedure can be applied to conducting a systematic review of a field of research. We compare five datasets that focus on the research of literature-based discovery (LBD).

LBD aims to foster new discoveries based on existing studies published in the literature. In particular, LBD aims to reveal and make use of undiscovered public knowledge. Don Swanson's pioneering work in the 1980s has been a major source of inspiration (16-19), followed by a series of studies in collaboration with Neil Smalheiser (20-22), who recently reviewed the past, present, and future of LBD (23). Significant developments have also been made by other researchers along this generic framework, for example, (24-26). A relevant concept of literature-related discovery (LRD) was proposed later on to expand the scope of LBD with literature-assisted discovery (LAD), which refers to the involvement of researchers behind the corresponding literature (27). In this study, we focus on LBD and consider research topics as relevant as long as they appear in our datasets.

Query-based search is a commonly used practice, especially when applying a localism approach to science mapping. The baseline dataset in this study is resulted from a full text search on Dimensions. Two datasets are constructed based on a groundbreaking article for LBD by Don Swanson (16). Two other datasets are constructed based on a recently published review paper on the work of Don Swanson as the pioneer of LBD research (23). All five datasets used Dimensions as the primary source.

The development of a scientific field may go through various stages. For example, Shneider proposed a four-stage model to characterize how a scientific field may evolve, namely, identifying the problem, building tools and instruments, applying tools to the problem, and codifying lessons learned (28, 29). The application stage may also reveal unanticipated problems, which in turn may lead to a new line of research and form a new field of research. The LBD research may continue along the research directions set off by the pioneering studies in LBD. According to Shneider's four-stage model, new specialties of research may emerge. A systematic review of LBD should cover such new developments as well as the established ones that we are familiar with. To what extent can we capture the relevant literature with a simple query? What would we miss if we rely on the result of the conventional search strategy alone? What would be the strengths of variations of the incremental expansion strategy?

To address these questions, we introduce an intuitive visual analytic method for comparing distinct search strategies across the underlying research landscape. This method enables us to examine thematic areas covered or missed by datasets collected with specific strategies and thus characterize the strengths and weaknesses of a specific procedure.

The rest of the article is organized as follows to detail the study. We characterize existing science mapping approaches in terms globalism and localism, especially their strengths and weaknesses for conducting systematic reviews of relevant literature. We introduce a network expansion methodology – cascading citation expansion – that bridges the divide between globalism and localism and improves the overall quality of systematic reviews. We apply the methodology to a systematic review of the landscape of LBD research and demonstrate how five datasets obtained from different strategies differ in terms of their coverage. Major clusters of the field and unique thematic patterns identified by individual datasets are visualized in the context of all datasets combined. Finally, we discuss the results and practical implications on conducting systematic reviews.

# Mapping the Scientific Landscape

Current science mapping approaches can be characterized as global, local, and hybrid in terms of their intended scope and applications.

## *Globalism*

Global maps of science aim to present a holistic view of the literature of all scientific disciplines (12). Commonly used units of analysis in global maps of science include journals and to a much less extent articles. Clusters of journals are typically used to represent disciplines of science.

Having a global map of science handy has several advantages for developing an understanding of the structure of scientific knowledge. For example, the global map gives users a stable representation to work with. Dealing with a stable structure is easier than dealing with an unstable one. A stable global structure is also useful in organizing additional details in the form of an overlay, a layer of information that can be superimposed over the stable base map.

Global maps may reveal insights that may not be possible at a smaller scale. For example, in studies of structural variations caused by scholarly publications, detecting a potentially transformative link in a network representation of the literature relies on the extent to which the contexts of both ends of the link are adequately represented (30).

Global maps also provide a relatively stable organizing framework for the end users. Klavans and Boyack compared the structure of 20 global maps of science (31). They arrive at a consensus map generated from edges that occur in at least half of the input maps. Researchers have tapped into the broad coverage of such global maps of scientific literature and have routinely used such global maps as a base map so as to present more focused information in the form of overlays in context as shown in (13, 14, 32).

On the other hand, globalism has disadvantages that may substantially undermine its advantages. For example, the validity of the base map is likely to diminish over time as subsequent publications constantly alter the underlying knowledge structure (30). Meanwhile updating a global map of science is not a simple task (14). How often should the base map get updated? If it is updated too often, the base map may lose its value as a relatively stable organizing framework. In contrast, if it is not updated for too long, then the map is increasingly out of date. Furthermore, as far as the frequency of update is concerned, there is little that end users can do because the tools and resources required to generate global maps are often not readily accessible to individual end users.

### *Localism*

To localism, the size of a science map is no longer the primary concern. Rather, the focus is on communicating the structure of a subject matter of interest at the level of scientific inquiry and scholarly communication, including analytic reasoning, hypothesis generation, and argumentation. Many studies of science belong to this category. Knowledge domain visualization, for example, focuses on knowledge domains as the unit of analysis (4, 33, 34). The concept of a knowledge domain is a problem-driven perspective. The boundary of a knowledge domain is not determined by the disciplinary boundaries nor by institutional boundaries. Rather, it is determined by the intrinsic relevance to the central research questions. A major practical challenge is the construction of the most representative body of the literature for a domain of interest based on an often much larger database.

Localized science maps are free from the need to maintain a stable structure. Instead, they are devoted to convey the latest structure of the state of the art and highlight the most intriguing aspects of the latest development in a field of study or intersections of a number of fields.

Commonly used resources of bibliographic data, especially citation data, range from the long-established sources such as the Web of Science and Scopus to relatively more recent additions such as Microsoft Academic Search and Dimensions. Collecting scientometric data using journals and subject categories alone are generally simple and common, but a major weakness of this strategy is the high risk of the inclusion of irrelevant articles and, perhaps more importantly, the exclusion of relevant ones. Multidisciplinary journals, for example, may require a different strategy to handle. Selecting a set of journals to adequately and accurately represent a knowledge domain can be challenging.

Query-based search is perhaps by far the most popular strategy for finding articles relevant to a topic of interest or a knowledge domain, which may include numerous interrelated topics. A query-based search process typically starts with a list of keywords or phrases provided by an end user. For example, a query of "*literature-based discovery*" OR "*undiscovered public knowledge*" could be a valid starting point to search for articles in the field of LBD.

Formulating a good query is a non-trivial task even for a domain expert because the quality of a query can be affected by several factors, including our current domain knowledge and our motivations of the search. Furthermore, if our goal is to identify emerging topics and trends in a research field, which is very likely when we conduct a systematic review of the field, then it would be a real challenging to express our targets precisely in advance. As a result, continuously refining queries over lessons learned from the performance of previous queries has been long recognized. Scatter/Gatherer (35), for example, has been influential on iteratively improving the quality of retrieved information based on feedback.

A profound challenge to query-based search is the detection of an implicit semantic connection, or a latent semantic relation. Search systems have utilized additional resources such as WordNet (36) and domain ontologies (37, 38) to enhance users' original terms with their synonyms and/or closely related concepts. With techniques such as latent semantic indexing (39) and more recent advances in distributional semantic learning (40), the relevance of an article to a topic can be established in terms of its distributional properties of language use. In other words, the semantic similarity of an article can be detected without an explicit presence of any keywords from users' original query. This is known as the distributional hypothesis: linguistic items with similar distributions have similar meanings.

### *Bridging the Gap*

Globalism and localism have been largely considered incompatible in the past, which prevent us from seeing how one may draw the strengths from both of them and reduced their weaknesses simultaneously. In this article, we conceptualize a unifying framework that accommodates both globalism and localism as special cases on a consistent and continuous spectrum by way of expansionism. Under the new conceptualization, the differences between a local map and a global map are reduced to two of the many possible states of the representation of the underlying scientific knowledge. Neither of them is seen as a static representation. Instead, they are states of an expansion process. Prior to the beginning of the expansion process, the state corresponds to the coverage of a localism map. The expansion process progresses by adding more publications into consideration and forms a wide variety of states that cover more than their corresponding local maps as starting points but less than the global map as the end point. In principle, the expansion process reaches its end point when its realm has reached and included all the articles ever published regardless disciplines, languages, and types. The end point would resemble to Memex - Vannevar Bush's visionary device proposed in 1945 (41), and it would be a superset of some of today's global maps of science, including those formed based on leading sources of scholarly publications such as the Web of Science, Scopus, Dimensions, and their possible combinations.

Table 1 summarizes the major advantages and weaknesses of globalism and localism along with potential benefits that the conceptualized incremental expansion may bring to us.

**Table 1. Comparisons between globalism and localism approaches.**

| Criteria for assessing the state of the art | Globalism | Localism | Incremental Expansion |
|---:|:---:|:---:|:---:|
| Accessible context | High | Low | Increased |
| Coverage, Diversity, Recall | High | Low | Increased |
| Redundancy | High | Low | Reduced |
| Structural stability | High | Low | Increased |
| Precision | Low | High | Increased |
| Sensitive to structural change | Low | High | Increased |

There may be multiple pathways connecting localism and globalism approaches. In this article, we demonstrate the flexibility and extensibility of a particular type of incremental expansion – cascading citation expansion – by applying this methodology to a few common scenarios in research in relation to a field of research of own interest, i.e. literature-based discovery (LBD).

## Cascading Citation Expansion

Citation indexing was originally proposed by Eugene Garfield to tackle the information retrieval problem in the context of scientific literature (42). The idea of citation indexing is straightforward. Researchers cite published articles in their own publications. The nature of a citation may vary widely as many researchers have documented (43). Regardless of the arguments for or against whether citations serve as a reliable or an accurate estimate of the scholarly impact of the work in question, an instance of a citation from one article to another provides justifiable evidence that some meaningful connections may exist between the two articles. In other words, we have a reason to believe that both articles are potentially relevant if one of them is considered relevant. A unique advantage of a citation-based search method is that it frees us from having to specify a potentially relevant topic in our initial query so as to reduce the risk of missing important relevant topics that we may not be aware of.

We often encounter situations in which we have found a small set of highly relevant articles and yet they are still not fully satisfactory for some reasons. For example, if we were new to a topic of interest and all we have to start with is a systematic review of the topic published many years ago, how can we bring our knowledge up to date? If what we have is a newly published review written by a domain expert, how can we expand the scope of interest from this particular review to a broader context? Another common scenario is when we want to construct a relatively comprehensive landscape of the existing literature on a target topic, how can we generate a dataset that is comprehensive enough but also contains the least amount of less relevant articles.

When we encounter these situations, switching to a completely new search altogether may not be effective. Whenever possible, we should be able to make use the search results we have obtained so far rather than throw them away. The real challenge is how we can find a way out from the current situation smoothly. Scatter/Gather (35) was a dynamic clustering strategy proposed in mid-1990s by allowing users to re-focus on query-specific relevant documents as opposed to query-independent clusters in browsing search results. When dealing with scientific publications, citation indexing provides a valuable mechanism to retrieve documents that are potentially relevant but may have a low similarity ranking due to the so-called vocabulary mismatch problem. From the citation indexing point of view, articles that cite any articles in the initial result set are good candidates for further consideration. Thus, an incremental expansion strategy can be built based on these insights so as to uncover additional relevant articles.

An incremental expansion process consists of one or more sub-processes that can be performed in parallel or in sequence. A sub-process consists of a sequence of expansion steps. Each expansion step is applied to a subset S of the scientific literature and produces a superset S+ of the subset as the result. The increased portion S+-S consists of publications that are the most relevant ones given S.

A useful type of incremental expansion is the cascading citation expansion method that we introduced recently (44). The idea of cascading citation expansion is not new, but it has been generally inaccessible to the research community except a small number of groups.

Cascading citation expansion is a search process that collects relevant articles based on citation links between a set of articles that one has already retrieved and the rest of the scientific literature (44). The process runs in successions in terms of the number of citation links from the original set of articles to the articles generated in the last round of expansion. In principle, an expansion process is capable of covering an arbitrarily large area of the entire scientific literature if one strategically selects the initial set of seed articles.

In order to enable cascading citation expansion, it is necessary to maintain a constant programmatic access to the entire collection of scientific publications. This bottleneck has been a major obstacle until very recently as Digital Science's Dimensions started to support the API access to their collection of over 98 million publications. A cascading citation expansion process may contain two sub-processes of forward and backward citation expansions with reference to an initial set of publications as a starting point. The initial set corresponds to the construction of a local map. Each step of expansion enlarges the accumulated coverage of the expansion process.

One may envisage two plausible scenarios if the expansion process keeps running continuously. If the entire universe of all the scientific publications is completely reachable from one point to another, then the expansion process will eventually visit every publication. In other words, the expansion process may start from a local map and expands its coverage to the entire collection, which is equivalent to the global map of scientific publications. Another scenario is that the entire universe is not always reachable from one point to another. Instead, the scholarly universe may consist of multiple galaxies of publications. Publications within the same galaxy would be reachable via both forward and backward citation expansions, whereas publications from different galaxies would not be reachable from one another. Consequently, if the initial set is fully contained within a single scholarly galaxy, then the expansion process can only exhaust the same galaxy but will not be able to go beyond it. Pragmatically speaking, the question is whether this is sufficient as far as the expansion is concerned. The exhausted galaxy is self-contained from the point of view of the relevance based on citation chains. A more robust strategy is to construct the initial set based on a query-based search and then utilize citation chains to expand the coverage. This strategy has the advantage of being able to initiate multiple expansion sub-processes simultaneously in multiple galaxies. The second scenario is more generic and includes the second one as a special case.

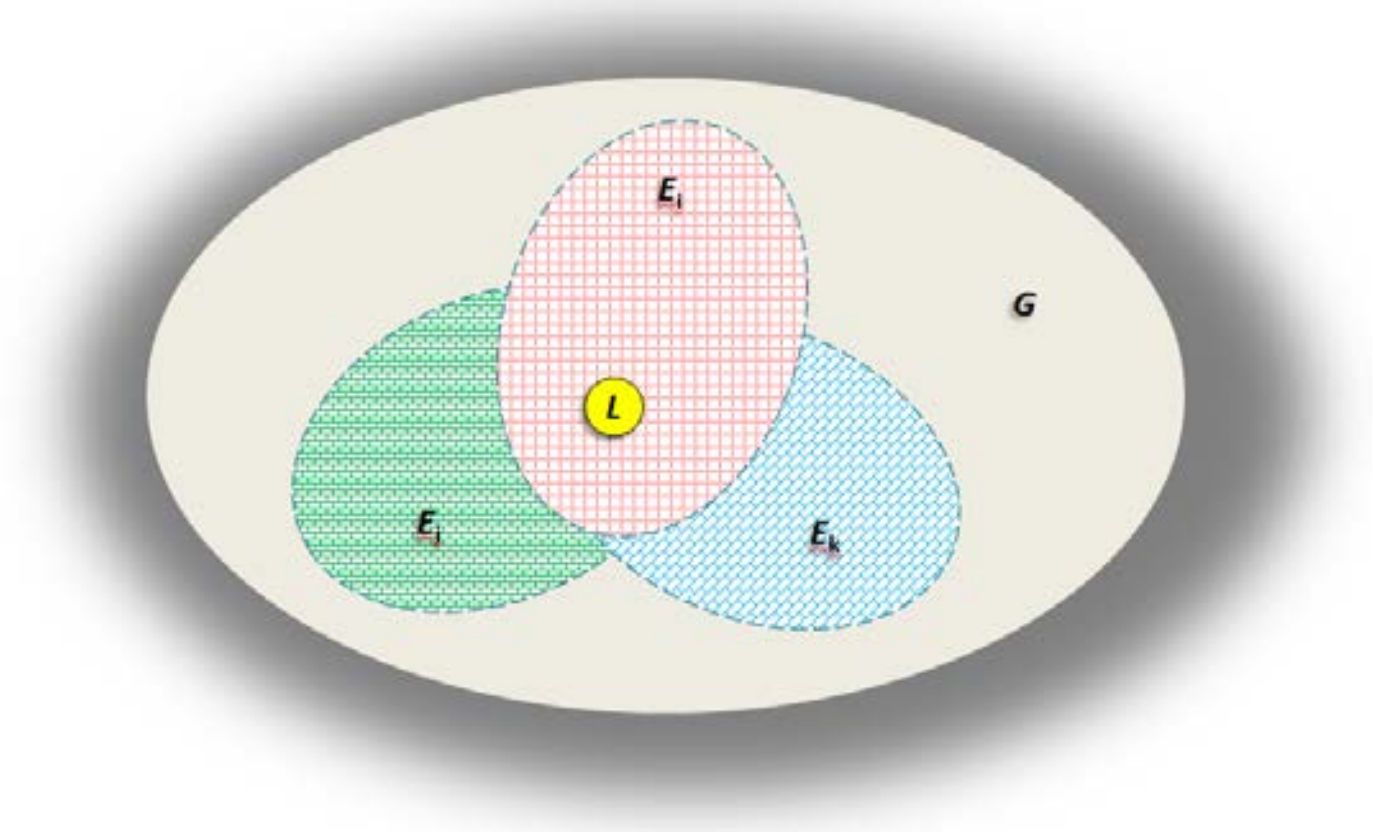

**Figure 1. A series of incremental expansions $E_i$, $E_j$, …, $E_k$ can arbitrarily reduce the difference between a local map (L) and a global map (G).**

## *Use Scenarios*

In this study, we consider two most common scenarios in research. In the first scenario, we have a well-known classic work to start with and one would typically like to find subsequently published studies that have built on the classic work. In the literature-based discovery research, Don Swanson's 1986 article on fish oil and Raynaud's syndrome (16) is a good example of such classic publications. Since its publication in 1986, what are the major milestones of the development over the past 30 years? What are the newly developed major topics ever since? What are the hottest and the most far-reaching topics in recent years? Given the original starting point of the 30-year development, what are the most unexpected topics emerged along the evolutionary paths?

In the second scenario, what we have is a recently published article and the reader would like to find precedent studies as well as other studies that are available in the scientific literature. An example at hand is a 2017 review article written by Neil Smalheiser (23), who has co-authored with Don Swanson on several landmark studies in the development of literature-based discovery. Smalheiser cited 71 references in his review. From the simple full text search for literature-based discovery, we know there are at least 1,777 published articles in the pool. What are the contexts from which Smalheiser as a domain expert selected the 71 references, which are less than 4% of the full text search results? What are the other major studies that are closely connected to the 71 selected references but not picked by the authoritative domain expert? What are the topic areas that are not covered by Smalheiser's review but nevertheless relevant to a broader context of the research field?

Pragmatically, if we were to rely on the simple full text search alone, how much would we miss? Perhaps more importantly, are there topics that we might have missed completely? What would be an optimal search strategy that not only adequately captures the essence of the development of the field as a whole but also does in the most efficient way? Borrowing the terminology from information retrieval, an optimal search strategy would maximize the recall and the precision at the same time.

Cascading citation expansion functions are implemented in CiteSpace based on the Dimensions' API. The expansion process starts with an initial search query in DSL, which is Dimensions' search language. Users who are familiar with SQL should be able to recognize the resemblance immediately. The result of the initial query forms the initial set of articles. In fact, in addition to publications, one can retrieve grants, patents, and clinical trials from Dimensions. In this study, we concentrate on publications.

The initial set can be a singleton set as well as a set of multiple articles. Swanson's 1986 article on fish oil and Raynaud's syndrome is an example of a singleton set, whereas the 1,777 full text search results could be an example of a set of multiple entities. The initial set serves as the starting point of a cascading citation expansion, which may consist of forward citation expansion, backward citation expansion, or both. A forward citation expansion retrieves articles that cited the current collection of articles. The newly retrieved articles will be added to the current collection of articles. We refer to each expansion step as a generation of an expansion process. If a forward expansion is completed at the 3rd step, then the entire process is called a 3-generation forward expansion. In contrast, a backward citation expansion retrieves articles that are cited by articles in the current collection. In other words, a backward citation expansion retrieves cited references and references cited by references. Similarly, we can define a k-generation backward citation expansion. More generically, we can define a m-generation backward and n-generation forward citation expansion, which is in fact what we did with Smalheiser's review.

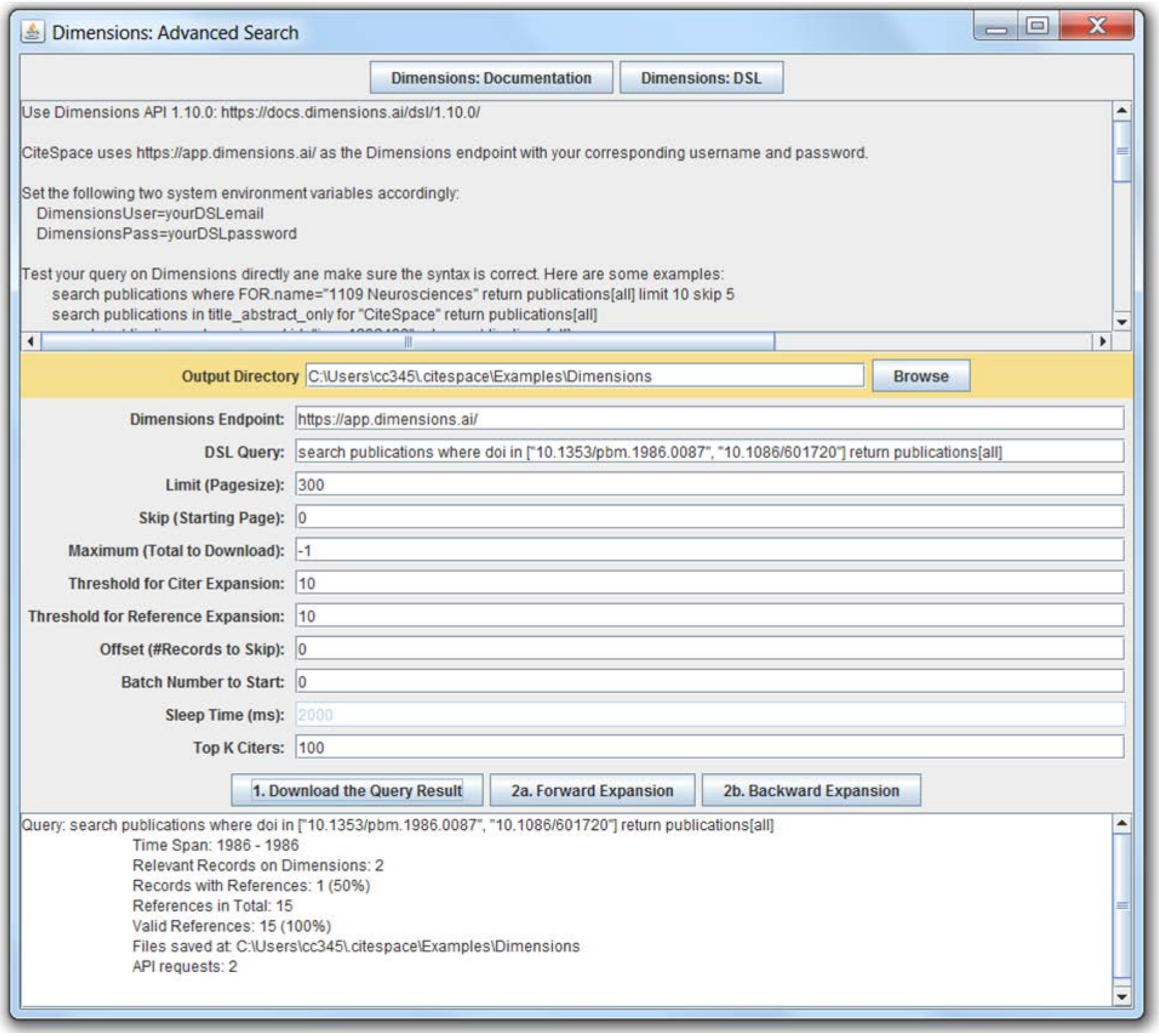

Figure 2. Citation expansion functions 2a and 2b are enabled after the initial search.

Figure 2 shows the user interface for cascading citation expansion in CiteSpace (44). The initial query is specified in the DSL query field. Users can apply various filters to the expansion process. For example, a threshold for citer expansion of 10 would allow the expansion go through a forward citation expansion if an article is cited by a citer that itself has been cited by 10 or more articles. Similarly, a threshold for reference expansion of 10 would allow a backward expansion if an article cited a reference and the reference itself has at least 10 citations. These filters provide users with a flexible trade-off option between concentrating on major citation paths with a reduced completion time versus retrieving articles comprehensively with a much longer completion time. Since the distribution of citations of articles follows power law, a comprehensive expansion process may become too long to be viable for a daily use of these functions. The DSL query shown in Figure 2 searched for Swanson's two articles published in 1986, namely, the fish oil and Raynaud' syndrome article 1986a (16) and the undiscovered public knowledge article 1986b (17). Swanson 1986a has 421 citations on Dimensions, whereas Swanson 1986b has 157 citations. The report at the bottom of the screen indicates that only one of the two articles contains references on Dimensions. Swanson 1986a is the one found with 25 references. Because of the citation threshold filters, 15 of the 25 references are qualified and retrieved. The fish oil article along with the 15 qualified references form the initial set for subsequent citation expansions.

### *Constructing Five Datasets of Literature-Based Discovery*

To demonstrate the flexibility and extensibility of the incremental expansion approach, we take the literature-based discovery (LBD) research as the domain of interest. We choose this topic because several reasons: 1) we are familiar with the early development of the domain in terms of the groundbreaking publications in this context, 2) we are aware of a recent review written by one of the pioneer researchers and we would like to set it in a broader context, and 3) we would like to take this opportunity to demonstrate how one can apply the methodology to a visual exploration of the relevant literature and develop a good understanding of the state of the art. These reasons echo the common scenarios discussed earlier.

In this study, we address three use scenarios with five datasets of literature-based discovery (Table 2). F represents a query-based search, the first scenario. $S_3$ and $S_5$ represent the second scenario in which all we have to start with is a groundbreaking article of the field and we wish to identify a representative dataset of the literature. $N_F$ and $N_B$ represent the third scenario in which all we have to start with is a recently published review article of the field and we would like to find the body of the literature that the review article is built on. All the datasets are retrieved from Dimensions through CiteSpace.

Table 2. Five datasets on literature-based discovery.

| **Set** | **Description** | **Records** | **Records with Abstracts on PubMed** |
|---|---|---|---|
| F | Fulltext search on Dimensions for "*literature-based discovery*" OR "*undiscovered public knowledge*" (Data of Search: 3/4/2019) | 1,777 | 431 |
| $S_3$ | 3-generation forward expansion from a seed article by Swanson (16)<br>Dimensions:pub.1053548428<br>Times Cited: 407<br>Citing Article and Reference Thresholds: 10, 10 | 739 | 516 |
| $S_5$ | 5-generation forward expansion of Swanson's pioneering work (16)<br>Citing Article and Reference Thresholds: 20, 20 | 43,703 | 22,918 |
| $N_F$ | Forward expansion from a 2017 seed article by Smalheiser (23) | 73 | 59 |
| $N_B$ | Backward expansion from $N_F$ | 2,435 | 1,830 |
| | Combined | 46,756 | |

Figure 3 shows logarithmically transformed distributions of the five datasets. The distributions shown under the title are the original ones.

- The F dataset (in blue) is evenly distributed except a surprising peak in 2009, which turns out due to many articles from an encyclopedia. The number of articles each year ranges between 60 and 130.
- Both $S_5$ (red) and $S_3$ (orange) are forward expansions starting with Swanson's 1986 article on fish oil and Raynaud's syndrome (16). In $S_3$, the inclusion threshold was at least 10 citations, whereas it was 20 in $S_5$ so as to keep the total processing time down. The majority of the articles in $S_3$ appeared between 2006 and 2018. $S_3$ had the first peak in 1984. It didn't return to the same level for the next 10 years until it started to climb up from 2003 and reached the second peak in 2012. In contrast, the distribution of the more extensive forward expansion $S_5$ shows a steady increase all the way over time.

- $N_F$, a forward expansion from Smalheiser's review of Swanson's work (23), includes articles that cited the references used in Smalheiser's review. Its distribution steadily increased between 1986 and 2017 with a rise of 5 over the 31-year span till 2016.
- $N_B$, a backward expansion from the set $N_F$, contains articles that are cited by $N_F$. The earliest article in $N_B$ was published in 1936. A noticeable hump from 1983 followed by a valley around early 1990s. Two peaks appeared in 2006 and 2011, respectively.

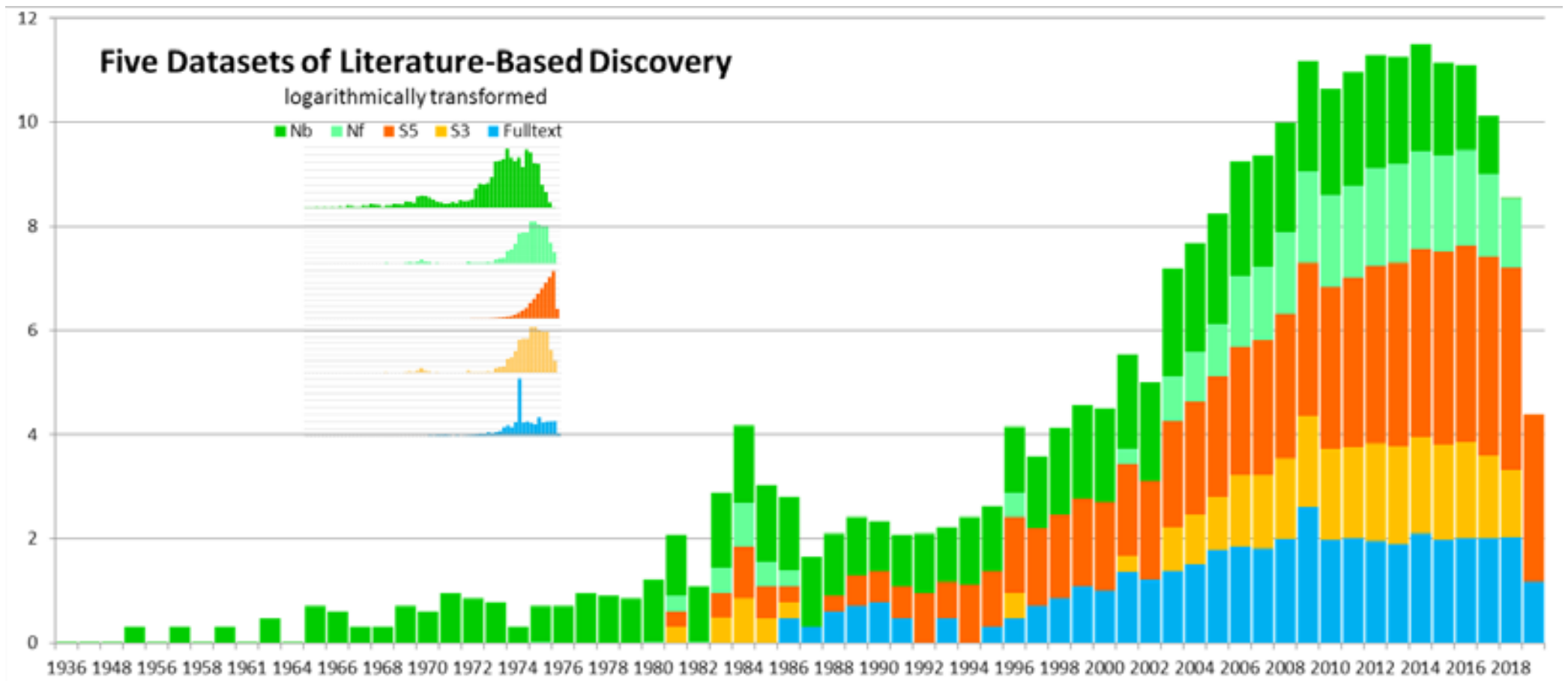

Figure 3: Distributions of articles by year in the five datasets. Full text search (F - blue) has a relatively even distribution over the years. Backward expansion of the recent review ($N_b$ - green) has the longest coverage.

Table 3 summarized how much these datasets overlap, measured as the ratio of articles in common to their set union. The five datasets together contain 46,756 unique articles published between 1936 and 2019. S5 represents 93.47% of the combined set of articles. NB represents 5.21%. F represents 3.80%.

**Table 3. Five individual datasets and a combined set.**

| | ***Combined*** | ***F*** | $S_3$ | $S_5$ | $N_F$ | $N_B$ |
|---|---|---|---|---|---|---|
| **Range** | 1936-2019 | 1936-2019 | 1975-2019 | 1975-2019 | 1986-2019 | 1956-2019 |
| **Articles** | 46,756 | 1,777 | 739 | 43,703 | 73 | 2,435 |
| ***Combined*** | 100.00 | 3.80 | 1.58 | 93.47 | 0.16 | 5.21 |
| ***F*** | 3.80 | 100.00 | 8.53 | 1.57 | 65.75 | 7.72 |
| $S_3$ | 1.58 | 3.55 | 100.00 | 1.65 | 21.92 | 4.15 |
| $S_5$ | 93.47 | 38.55 | 97.70 | 100.00 | 72.60 | 15.44 |
| $N_F$ | 0.16 | 2.70 | 2.17 | 0.12 | 100.00 | 5.82 |
| $N_B$ | 5.21 | 8.93 | 13.67 | 0.86 | 100.00 | 100.00 |

Table 4 summarizes properties of networks constructed from each of the five datasets as well as the combined set. Networks of all individual datasets are generated in CiteSpace with the same configuration, namely the link-to-node ratio=4, look back years=10, at least cited once, and ranked as top 100 most cited articles per year. The network of the combined set used the same configuration except the inclusion of top 300 most cited articles per year. For each network, the largest connected component (LCC) is shown along with the percentage in the entire network.

Both modularity and silhouette scores of a network are associated with how the network is decomposed into groups of nodes, often known as clusters. The modularity score of a network reflects the clarity of the network structure at the level of decomposed clusters. The silhouette score of a cluster measures the homogeneity of its members. A network with a high modularity and a high average of silhouette scores would be desirable.

As shown in Table 4, the combined network consists of 14,743 nodes and 79,170 links. Its LLC has 56% of the nodes in the entire network (8,352). Its modularity of 0.94 is very high, but the average silhouette of 0.13 is low. LCC is the largest connected component of a network. The LCC of forward expansion ($N_F$) contains 87% of the entire network, which is the highest among the five datasets. The LCC of backward expansion ($N_B$) contains 44% of the entire network, which is the lowest. The LCC of the full text search (F) contains 69% of the network.

**Table 4. Properties of five network overlays.**

| | ***Combined*** | ***F*** | $S_3$ | $S_5$ | $N_F$ | $N_B$ |
|---|---|---|---|---|---|---|
| **Nodes** | 14,743 | 3,435 | 2,972 | 3,969 | 1,903 | 10,821 |
| **Links** | 79,170 | 16,138 | 15,264 | 22,607 | 8,620 | 51,405 |
| **LCC (%)** | 8,352 (56) | 2,385(69) | 2,324(78) | 2,822(71) | 1,665(87) | 4,829(44) |
| **Modularity** | 0.94 | 0.82 | 0.79 | 0.91 | 0.87 | 0.95 |
| **Silhouette** | 0.13 | 0.09 | 0.80 | 0.41 | 0.29 | 0.14 |

The modularity of a network measures the clarity of the network structure in terms of how well the entire network can be naturally divided into clusters such that nodes within the same cluster are tightly coupled, whereas nodes in different clusters are loosely coupled. The higher the modularity is, the easier to find such a division. The silhouette score of a network measures the average homogeneity of derived clusters (45). The higher the average silhouette score is, the more meaningful a group is in terms of a cluster. $N_B$ has the highest modularity of 0.95 and $S_5$ has the second highest modularity of 0.91. $S_3$ has the highest average silhouette score of 0.80, followed by $S_5$. We speculate that the lower threshold used in $S_3$ might contribute to its higher silhouette average because $S_3$ would retain a finer granularity of the citation linkage than $S_5$. The full text search has the modularity of 0.82, which is high, but its silhouette value of 0.09 is the lowest. In terms of modularity and silhouette scores, $S_3$ and $S_5$ are potentially strong candidates to serve as the basis of a systematic review.

## Literature-Based Discovery

In this section, we visualize the thematic landscape of the field of literature-based discovery from multiple perspectives of the five datasets. We will start with the full text search results and then cascading citation expansions.

### *Full Text Search*

The query for the full text search on Dimensions consists of 'literature-based discovery' and 'undiscovered public knowledge.' The phrase 'literature-based discovery' is commonly used as the name of the research field. The phrase 'undiscovered public knowledge' appears in the titles of two Swanson's publications in 1986. One is entitled "Fish oil, Raynaud's Syndrome (16), and Undiscovered Public

Knowledge” in Perspectives in Biology and Medicine and the other is “Undiscovered Public Knowledge” in Library Quarterly (17).

The full text search found 1,777 records. Dimensions’ export center supports the export of up to 50,000 records to a file in a CSV format for CiteSpace (3, 33). Publication records returned from Dimensions do not include abstracts. We were able to match 431 of them in PubMed and retrieved corresponding abstracts. Abstracts may be utilized for further exploration at the later stage of the analysis.

Figure 4 shows an overview map of the research field based on the 1,777 publications retrieved from the full text search. The color of a link indicates the earliest year when two publications were co-cited for the first time in the dataset. In this visualization, the earliest work appeared on the left-hand side, whereas the most recent ones appeared on the right-hand side, although CiteSpace does not utilize any particular layout mechanisms to orient the visualization in this way. The network is decomposed into clusters of references that are tightly coupled by how often they are co-cited. Clusters are numbered in the descending order of their size. The largest one is numbered as #0, followed by #1, and so on.

The largest cluster is #0 information retrieval, which appeared relatively early in the development of the field. In contrast, more recent clusters include #1 semantic predication, #3 computational drug (discovery), and #13 bibliometric study. Two of the most cited references in #3 are both published in 2014, whereas the most cited ones in #13 are from 2016 and 2017. Since our primary goal in this study is to evaluate the differences between various strategies of search, we will defer a more in-depth review of the field to later sections.

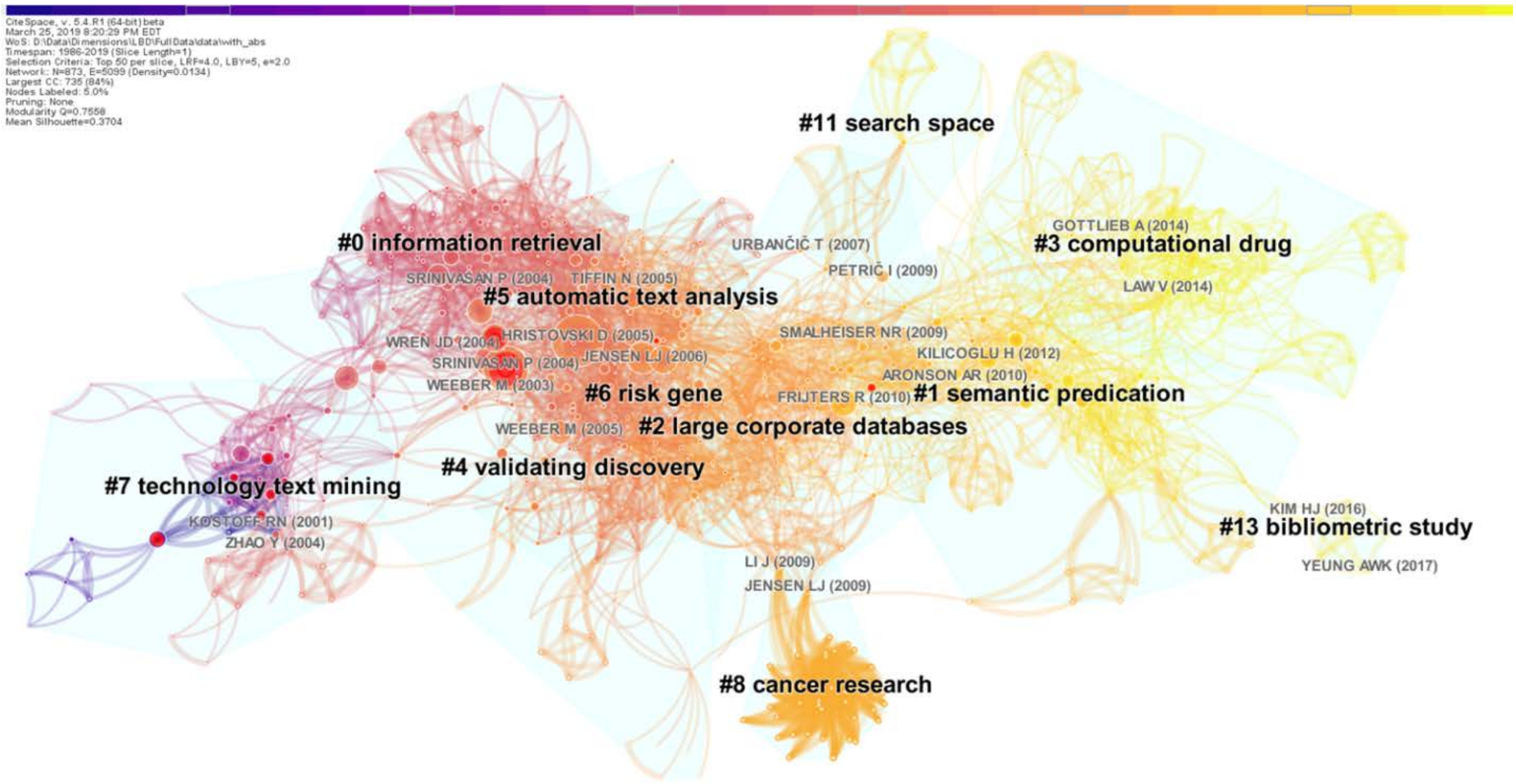

**Figure 4. A thematic overview of literature-based discovery based on a document co-citation analysis of the full text search result dataset. CiteSpace configuration: LRF=4, LBY=5, e=2.0, Top50 per slice.**

For the purpose of a systematic scientometric review, the general guideline is that one should maximize the coverage of the relevant literature so as to make the results as representative as possible. However, the initial simplicity may diminish rapidly if it becomes necessary to construct a complex query. Furthermore, no matter how complex a query one may conceive, such strategies cannot rule out the possibility of missing a potentially significant aspect. After all, the most valuable review of the literature would be the one that can bring us something new and draw our attention to something that we are not aware of or some potential relevance that we have not yet realized. Therefore, query-based search strategies alone tend to be limited by what we currently know and even by the vocabularies we may choose to describe the topic of our interest (39).

### *Full Text Search vs. Forward Citation Expansion*

According to Table 5, $S_5$ is a promising candidate to represent the knowledge domain of LBD. We will first demonstrate the differences between $S_5$ and the full text search.

In Figure 5, a document co-citation network derived from the full text search is superimposed over a base map derived from $S_5$, a 5-generation forward citation expansion from Swanson's article on fish oil and Raynaud's syndrome. Areas where they overlap are located in the lower right corner of the base map. In contrast, the large area located at the upper left corner is not covered by the full text search at all. Topics in the red box, for example, would be valuable for us to learn how these two retrieval methods differ.

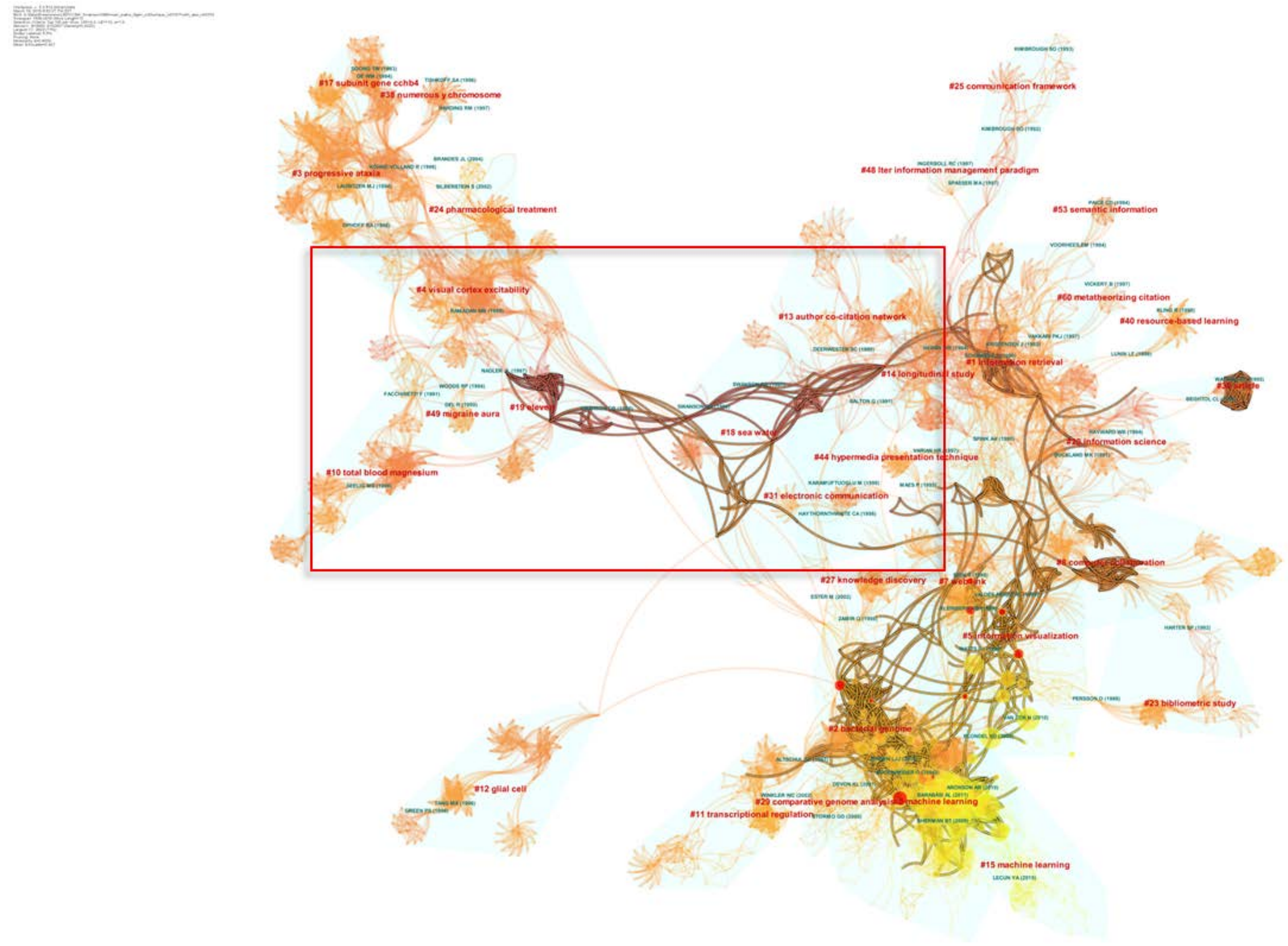

**Figure 5. A document co-citation network resulted from the full text search is superimposed over a document co-citation network derived from S5, a 5-generation forward citation expansion from Swanson 1986a (16).**

The area inside the red box in Figure 5 is shown Figure 6, which depicts how $S_5$ and the full text search differ. The solid lines in dark red indicate what the two datasets have in common, including three articles pointed by the dashed lines. The three articles are Swanson's articles. The one on the left is Swanson's 1988 article (46) on eleven neglected connections between migraine and magnesium. The leftmost cluster, #19 eleven, suggests that it is likely the footprint of the 1998 article. The one in the middle is the Swanson 1986 article (16), which is the seed article that the citation expansion process started with. The third one is Swanson's 1987 article that further consolidated his strategy of linking disparate bodies of literature for literature-based discovery. There are a few clusters on the left that are not covered by the full text search. However, these topics are still associated with migraine and magnesium, for example, migraine aura, visual cortex excitability, and total blood magnesium. It would be hard to generalize whether these extra clusters are definite gains for the expansion or definite losses for the full text search because it would depend on the role these extra clusters may play in a particular study. Nevertheless, we consider that having this option itself brings distinct advantages to researchers. Furthermore, since cascading citation expansion makes it explicit what could be missing or what might be added, one may benefit from an additional way to verify the scope of a field mapping.

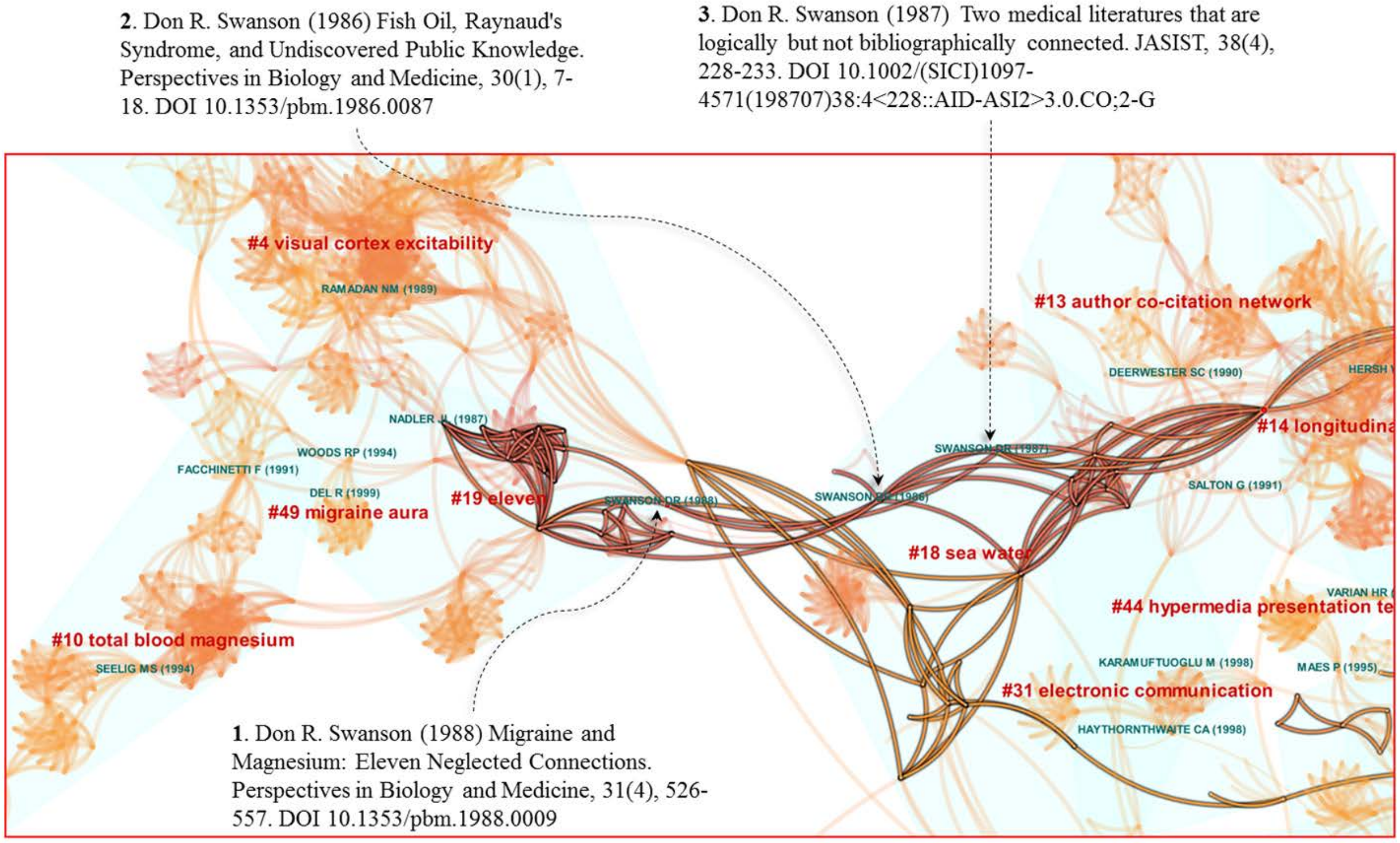

**Figure 6. How did the forward expansion ($S_5$) and the full text search (F) differ?**

## *The Structure of the Thematic Landscape*

In order to compare and contrast the coverage of individual datasets, we construct a base map with all the five datasets so that we can identify areas that are uniquely covered by a particular dataset (see Figure 7). The base map is divided into clusters. Any two datasets can be compared and contrast in terms of clusters covered by both of them, one of them, or none of them, just as we did with $S_5$ and full text search.

$S_5$ (Figure 7.c) and $N_B$ (Figure 7.f) covered some common areas as well as areas that are unique to themselves. Based on the colors, one can observe that $N_B$ comprehensively covered the earliest clusters located near the upper left corner of the base map, whereas the best coverage for clusters around the upper right corner was made by $S_5$. It is interesting to observe that, in essence, the full text network appears to be a superset of $S_3$ but a subset of $S_5$. Such interrelationships between individual datasets effectively improve our understanding of the role played by different areas in a broader context. It underlines core areas of the field regardless where an expansion process starts.

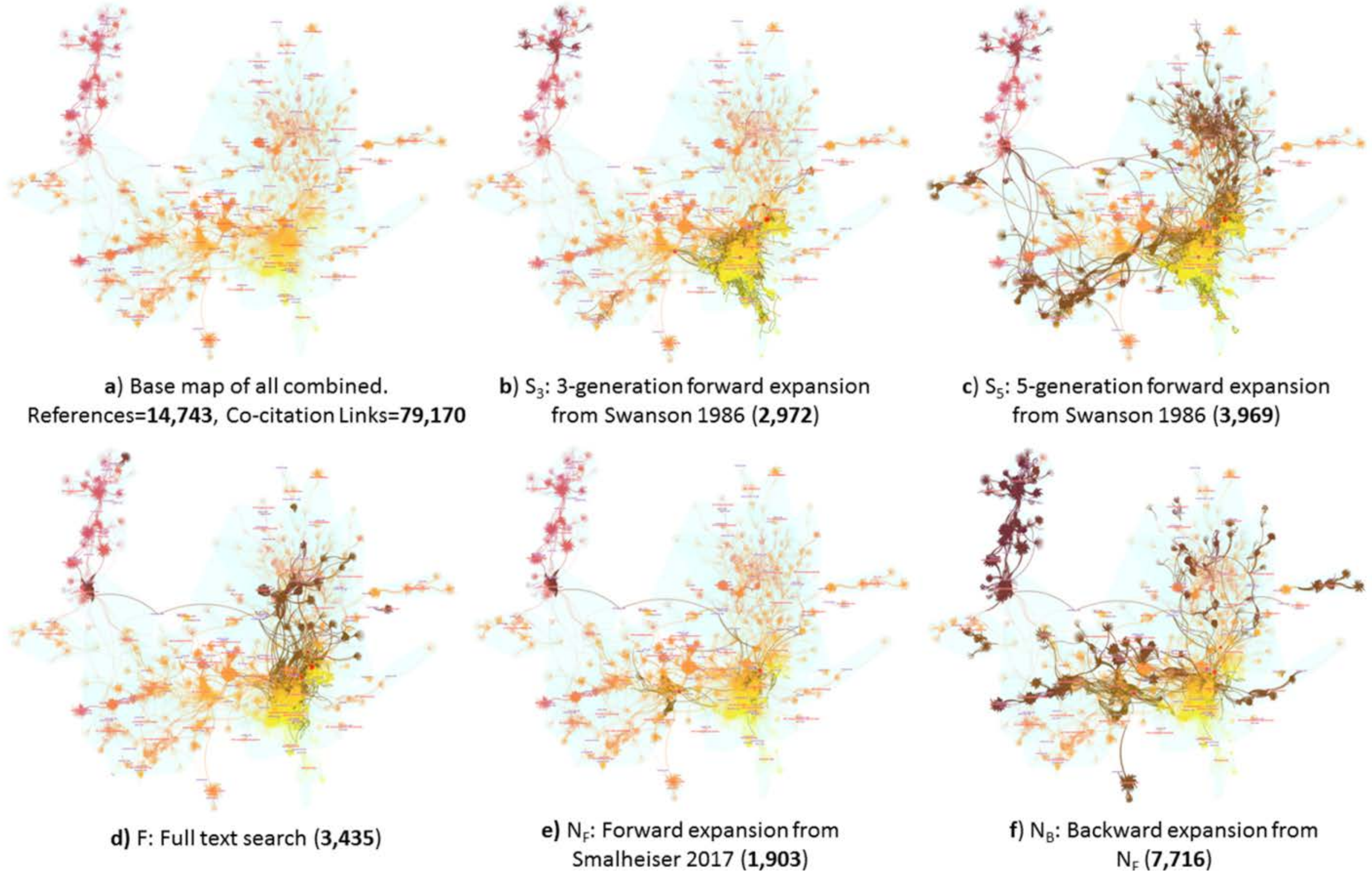

**Figure 7. The overlays of 5 individual datasets on the combined network as the base map.**

In Figure 8, we annotated areas that are in common between different individual approaches so that one can select an optimal strategy to achieve the same effect more efficiently. We use the notation $E_F^n(S)$ to indicate a forward expansion of a set S for *n* times. Similarly, $E_B^m(S)$ denotes a backward expansion of S for *m* times. Thus $N_B=E_BE_F(\{\text{Smalheiser 2017}\})$, which means 1-step forward expansion of a singleton set of Smalheiser's 2017 review, followed by a 1-step backward expansion of the result set of the previous step. The area surrounding cluster #2 Biomedical Literature is consistently covered by all the individual networks. We consider this as the core area of LBD. From the core area outwards, different approaches behave differently. The branch extending from the core to the left was mainly contributed by expansion processes with Samlheiser's review (23) as the seed ($N_F$ and $N_B$ in Figure 8.a), whereas the two branches from the core downwards were primarily resulted from the expansions starting with Swanson 1986a (16) ($S_3$ and $S_5$ in Figure 8.b). We will examine the specific themes of these branches shortly. According to the colors, the oldest clusters are located near the upper left region of the combined base map. Notably, #10 fish oil and #45 Raynaud's Phenomenon immediately remind us what the seed article (16) was about.

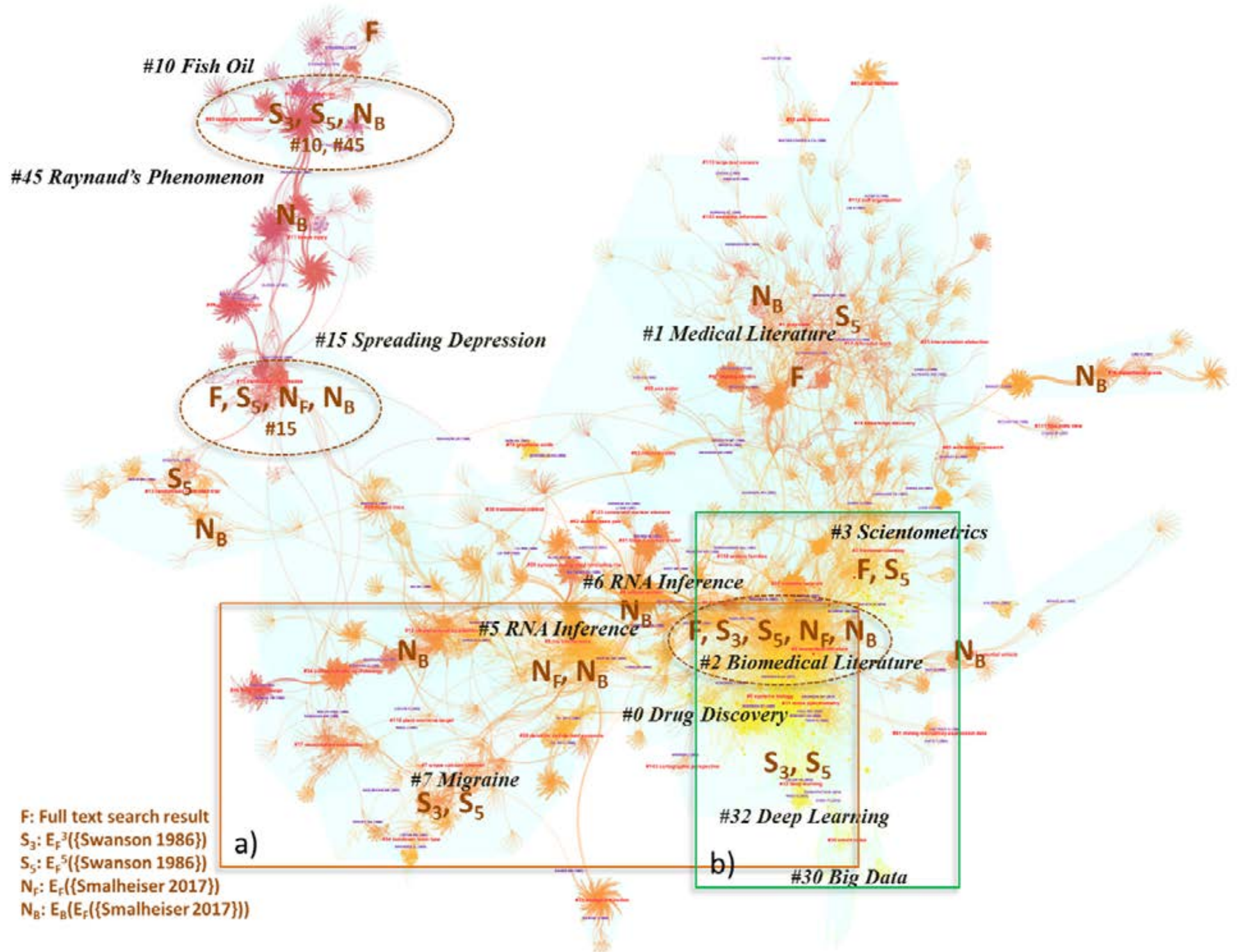

**Figure 8. Overlapping areas of the five datasets.**

Figure 9 contrasts the coverage of individual network overlays shown in Figure 8.a. The first row represents the coverage of the full text search F, which missed this region considerably, except the upper right corner. The second row illustrates the coverage of forward expansions. In particular, $S_3$ expanded further from the right towards the center and a small number of articles emerged near the far south, but the region as a whole remains under-represented. As it turns out, $S_5$ covered the region substantially, excluding the branch running horizontally near the top. The third row features the coverage of backward expansions. $N_F$ covered the common base of the upper and lower branches, whereas $N_B$ covered the entire upper branch. These patterns suggest that different search strategies may reach significantly different areas of the thematic landscape. More specifically, using the full text search alone may indeed miss a substantial part of the landscape and cascading citation expansion may approximate the underlying thematic landscape of a research field as detailed as the end user wishes.

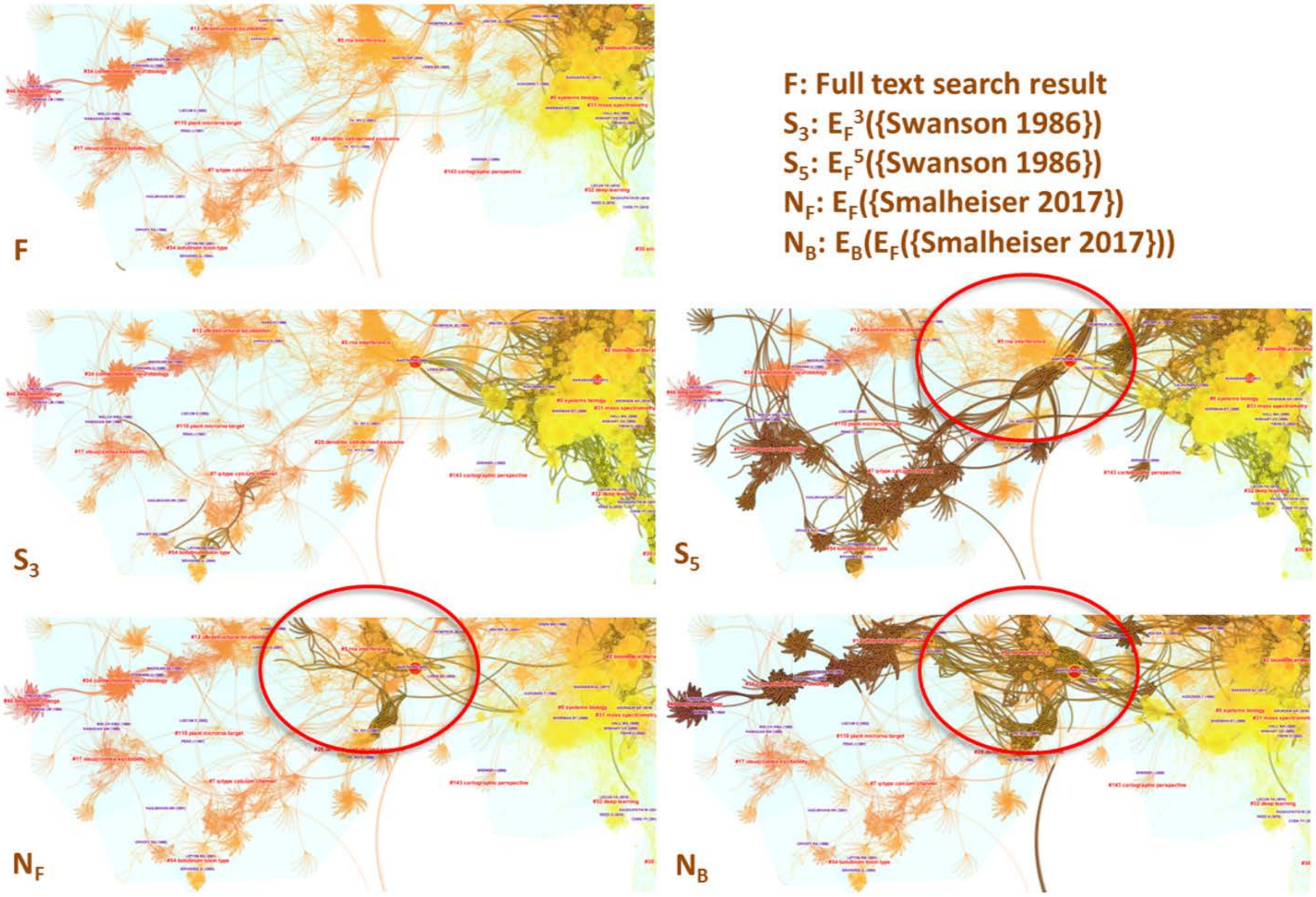

**Figure 9. The region of Figure 8.a.**

Figure 10 illustrates how three of the network overlays differ in the region of Figure 8.b. Several themes in this region are characterized by corresponding cluster labels in the middle image but omitted in the other two for clarity. Cluster #2 biomedical literature is featured in all the network overlays. Clusters #32 deep learning and #30 big data, located below cluster #2, are covered in full by $S_5$, but categorically missed by the full text search $F$ and the backward expansion $N_B$. Cluster #3 scientometrics, located above #2 biomedical literature, was almost absent from the network overlay of $N_B$, but it was covered extensively in the network overlays of $F$ and $S_5$. If we were to use the full text search alone in a systematic review of the field, the role of big data and deep learning is likely to be under-represented. The difference between $N_B$ and $S_5$ is even more interesting. It appears that a forward expansion from a pioneering study may lead to a more comprehensive coverage.

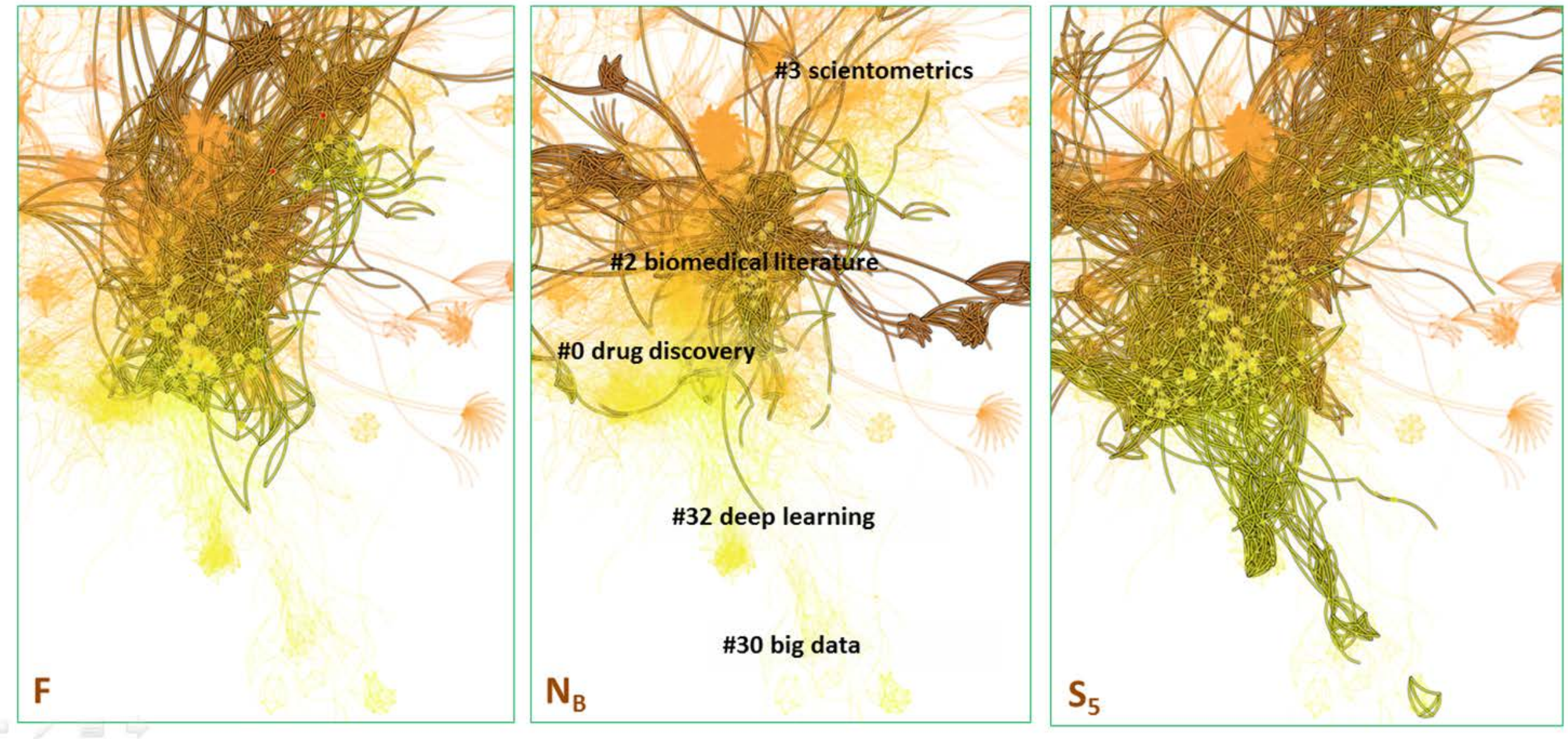

**Figure 10. The region of Figure 8.b.**

## *Major Themes*

A cluster of co-cited references in CiteSpace represents an underlying theme, as in the above example of #32 deep learning. In addition to the interrelationships between clusters, one may examine what a cluster is made of. There are several ways to explore a cluster in CiteSpace. The most convenient method is to use the cluster explorer. In this article, we demonstrate an iterative approach that can be repeatedly applied to a cluster at one level and its sub-clusters at the next level of granularity. The major advantage of this approach is that all the visual analytic functions in CiteSpace can be applied at multiple levels of analysis, in particular, including interactive exploration of a visualized network, examining citing and cited articles of a cluster, and organizing concepts in a hierarchical visualization, and detecting the burstness of various articles.

To understand the thematic structure of a cluster, we first present the cluster in the context of the entire thematic landscape. Then, we break down the cluster into its own clusters at the next level of granularity. This approach shares similarities with the principles behind scatter/gatherer in information retrieval (35). A top-level cluster, $\#c_i$, is further divided into several second-level clusters, or its sub-clusters denoted as $\#c_i^j$. Next, we check the main branches in a concept tree, generated from phrases found in relevant articles.

### *The Largest Cluster: Drug Discovery*

The largest cluster #0 is labeled as drug discovery. The top-5 of its sub-clusters $\#0^0$, …, $\#0^4$ are shown in the upper left quadrant of Figure 11, namely $\#0^0$ protein interaction and $\#0^1$ computational drug (discovery). The lower left diagram shows the context (shown in gray) of the top-5 clusters (shown in red). The concept tree in the upper right quadrant presents a hierarchy of the key phrases of the cluster on drug discovery, notably along two branches: drug repositioning and drug repurposing. The embedded table lists 5 articles that cited cluster #0 most extensively in terms of the number of cluster members cited. For example, the first article cited 30 references of the cluster and itself has been cited 214 times within the entire Dimensions' collection.

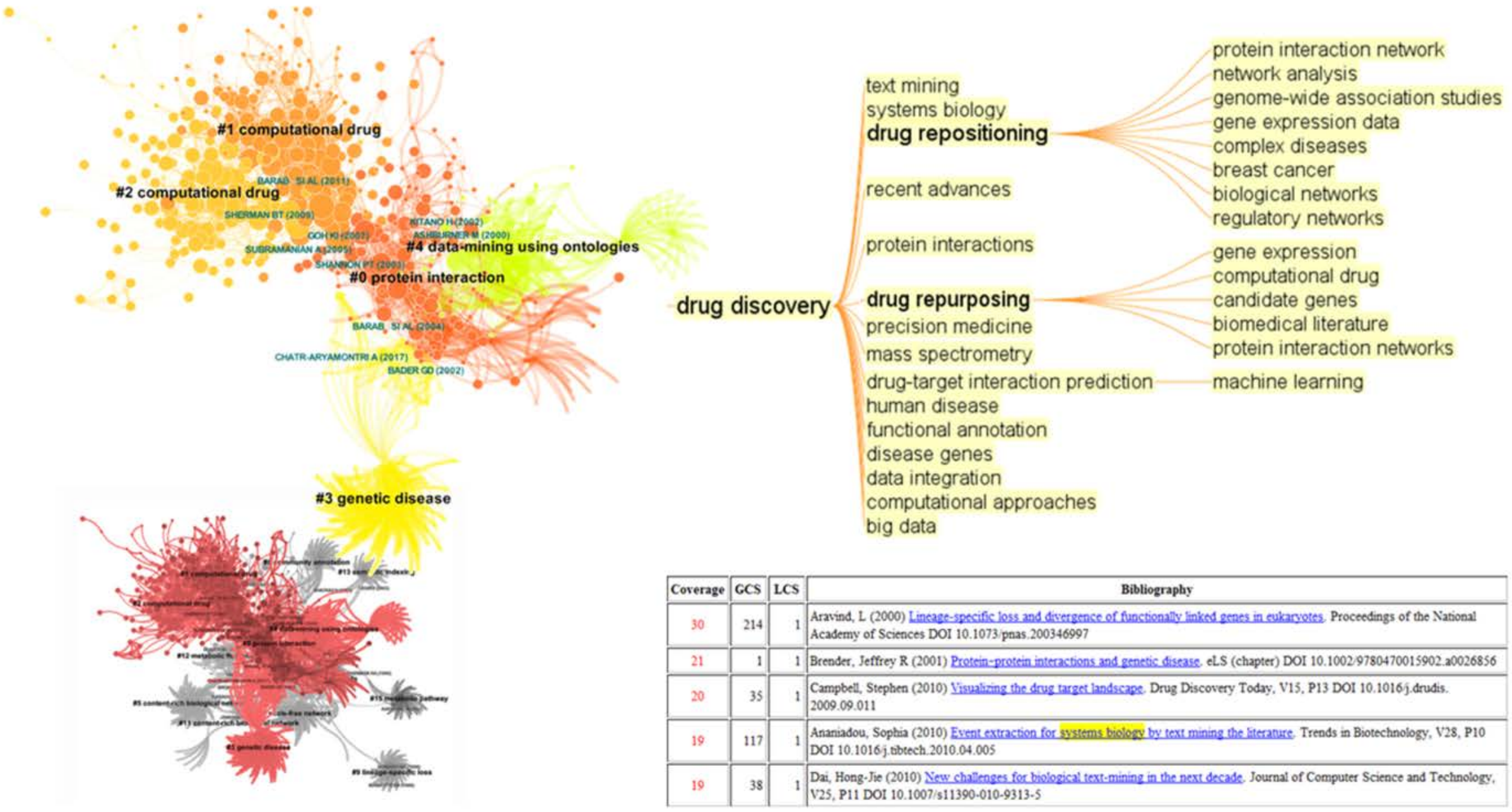

| Coverage | GCS | LCS | Bibliography |
|---|---|---|---|
| 30 | 214 | 1 | Aravind, L (2000) Lineage-specific loss and divergence of functionally linked genes in eukaryotes. Proceedings of the National Academy of Sciences DOI 10.1073/pnas.200346997 |
| 21 | 1 | 1 | Brender, Jeffrey R (2001) Protein-protein interactions and genetic disease. eLS (chapter) DOI 10.1002/9780470015902.a0026856 |
| 20 | 35 | 1 | Campbell, Stephen (2010) Visualizing the drug target landscape. Drug Discovery Today, V15, P13 DOI 10.1016/j.drudis.2009.09.011 |
| 19 | 117 | 1 | Ananiadou, Sophia (2010) Event extraction for systems biology by text mining the literature. Trends in Biotechnology, V28, P10 DOI 10.1016/j.tibtech.2010.04.005 |
| 19 | 38 | 1 | Dai, Hong-Jie (2010) New challenges for biological text-mining in the next decade. Journal of Computer Science and Technology, V25, P11 DOI 10.1007/s11390-010-9313-5 |

**Figure 11. The largest cluster in the combined network: #0 drug discovery.**

*The Second Largest Cluster: Medical Literature*

The second largest top-level cluster #1 is labeled as medical literature (Figure 12). The concept tree includes the term medical literature, which in turn has the term LBD systems as one of its children nodes. LBD is an abbreviation of literature-based discovery. The second branch on text mining is associated with its primary focus on biomedical literature. The second-level clusters $\#1^0$ spatial relationship and $\#1^1$ network-based retrieval model suggest that the top-level cluster #1 is concerned with LBD in the context of medical literature.

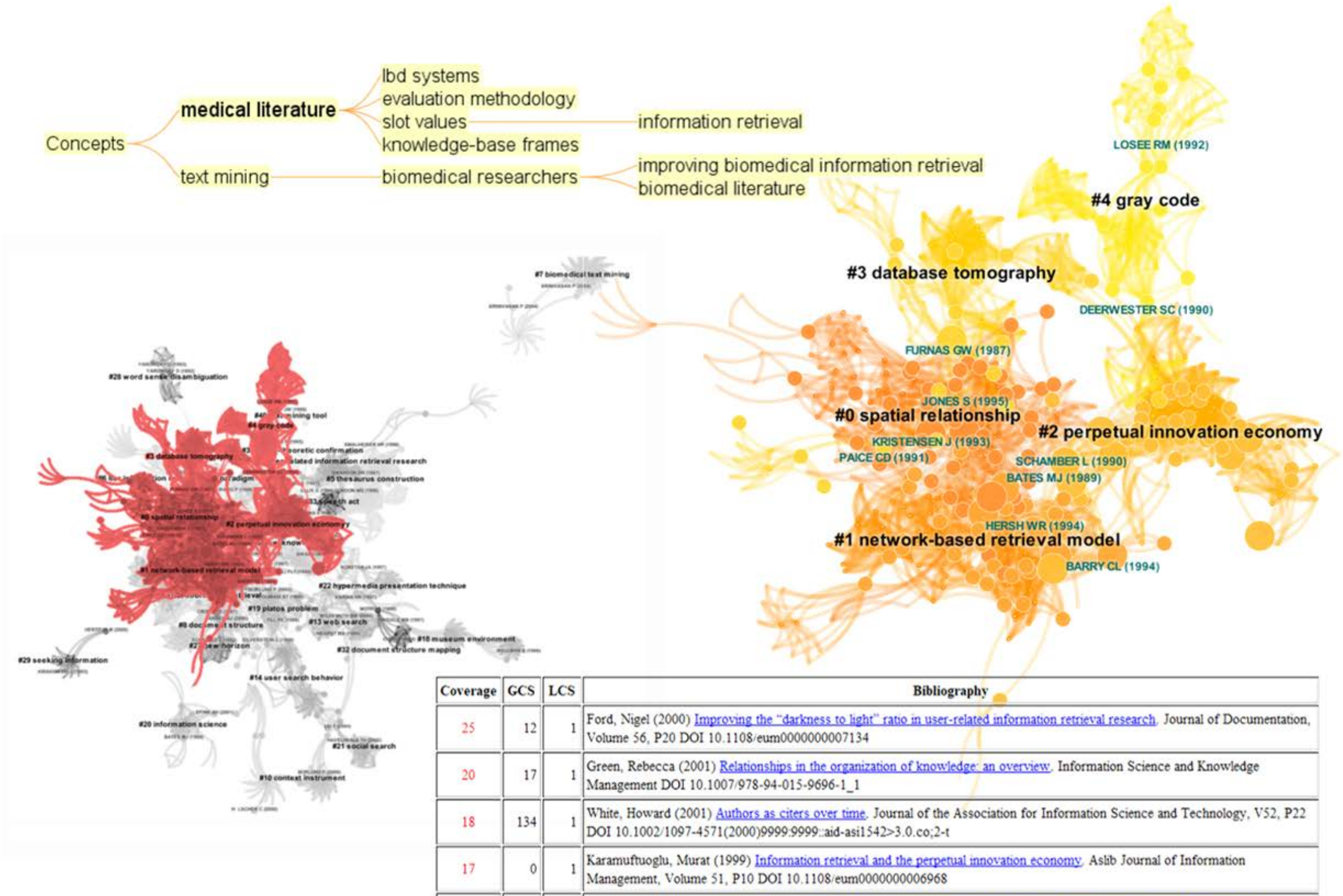

| Coverage | GCS | LCS | Bibliography |
|---|---|---|---|
| 25 | 12 | 1 | Ford, Nigel (2000) Improving the "darkness to light" ratio in user-related information retrieval research. Journal of Documentation, Volume 56, P20 DOI 10.1108/eum0000000007134 |
| 20 | 17 | 1 | Green, Rebecca (2001) Relationships in the organization of knowledge: an overview. Information Science and Knowledge Management DOI 10.1007/978-94-015-9696-1_1 |
| 18 | 134 | 1 | White, Howard (2001) Authors as citers over time. Journal of the Association for Information Science and Technology, V52, P22 DOI 10.1002/1097-4571(2000)9999:9999::aid-asi1542>3.0.co;2-t |
| 17 | 0 | 1 | Karamuftuoglu, Murat (1999) Information retrieval and the perpetual innovation economy. Aslib Journal of Information Management, Volume 51, P10 DOI 10.1108/eum0000000006968 |
| 16 | 15 | 1 | Marty, Paul F (1999) Museum informatics and collaborative technologies: the emerging socio-technological dimension of information science in museum environments. Journal of the Association for Information Science and Technology DOI 10.1002/(sici)1097-4571(1999)50:121083::aid-asi7>3.0.co;2-b |

**Figure 12. The second largest top-level cluster #1 medical literature.**

### *The Third Largest Cluster: Biomedical Literature*

The third largest top-level cluster is #2 biomedical literature (Figure 13). This is the thematic area consistently covered by all the individual network overlays, indicating a major aspect of the literature-based discovery research. Indeed, $\#2^0$ is labeled as literature-based discovery. The second-level clusters also feature $\#2^1$ entity recognition, $\#2^2$ big data, and $\#2^3$ genomic era. The concept tree highlights prominent themes of this area, notably big data, biomedical literature, and natural language processing. The concept of literature-based discovery appears as one of the children nodes of biomedical literature. It is also revealing to observe some of the longer paths in the concept tree, for example, biomedical literature – biomedical text mining – entity recognition – conditional random fields. The top-5 citing articles shown in the embedded table echo the common themes such as extracting information from the pharmacogenomics literature, mining biomedical literature, comparing patterns in bacterial genomes. Given that this area is consistently covered by all individual networks, these topics are among the most prominent ones in the landscape of literature-based discovery.

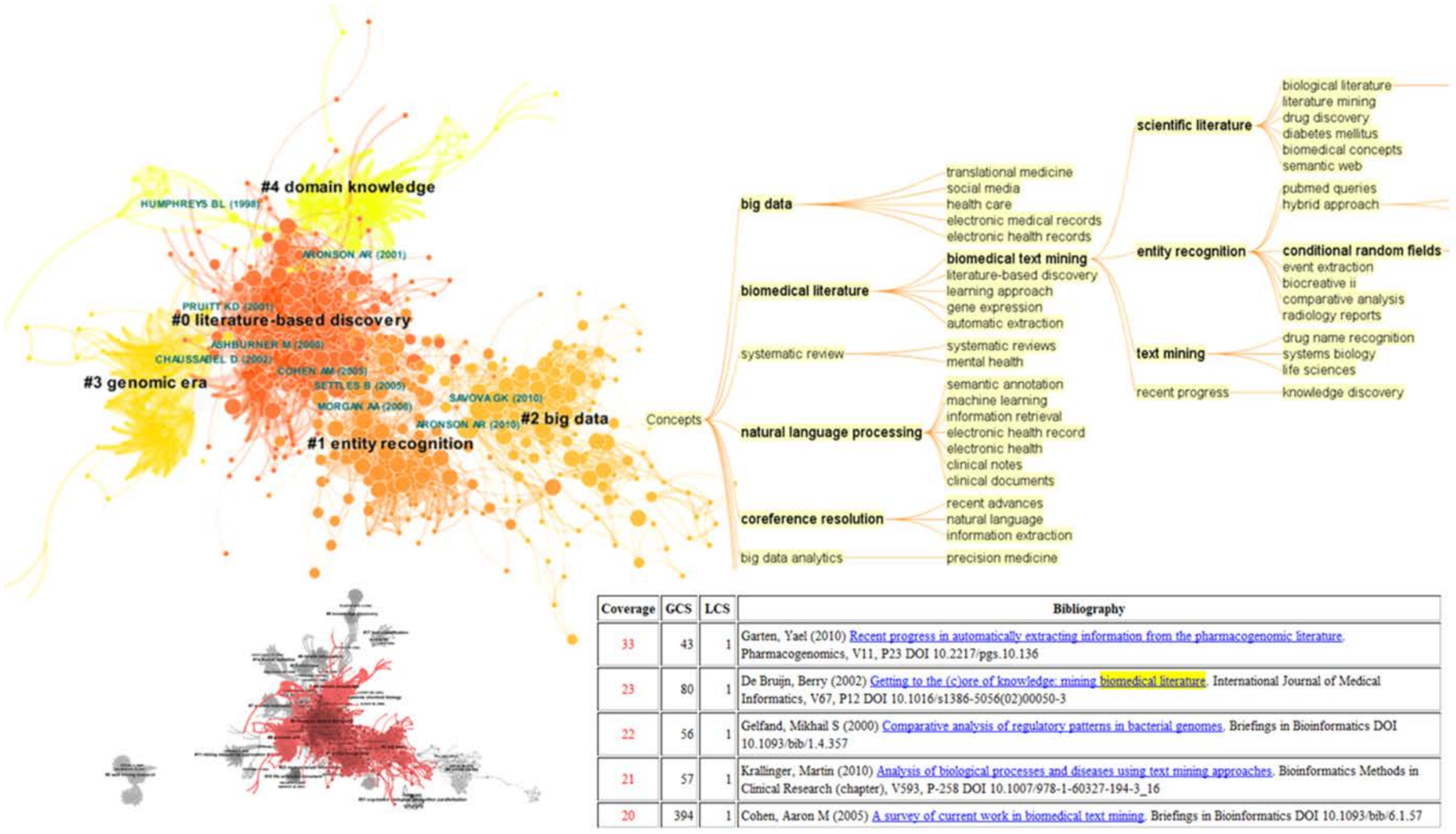

**Figure 13. The third largest top-level cluster #2.**

## *Special Themes*

All the five networks featured Cluster #2 Biomedical Literature to some extent. The second most popular cluster is Cluster #15, four of the five networks, covered this cluster either completely or partially.

As shown in Figure 8, four of the five individual networks considerably shared the top-level cluster #15, which is also linking the fish oil cluster (#10) and the Raynaud's phenomenon cluster (#45) and the much larger continent where the prominent clusters #0, #2, and a few others are located. Figure 14 reveals the structure of cluster #15 in terms of its top-5 second-level clusters.

The second-level clusters include $\#15^0$ magnesium / vascular head pain, $\#15^1$ migraine / hallucinogen ingestion, $\#15^2$ magnesium / cardiovascular disease, $\#15^3$ migraine / unnoticed connection, and $\#15^4$ association / acute migraine attack. These topics are clearly concentrated on migraine and magnesium. Cluster $\#15^3$ in particular contains two of Swanson's publications (16, 46). Given the strategic position of cluster #15 in the overall landscape of literature-based discovery, the role of migraine-magnesium connection in the evolution of the research field is worth further investigation.

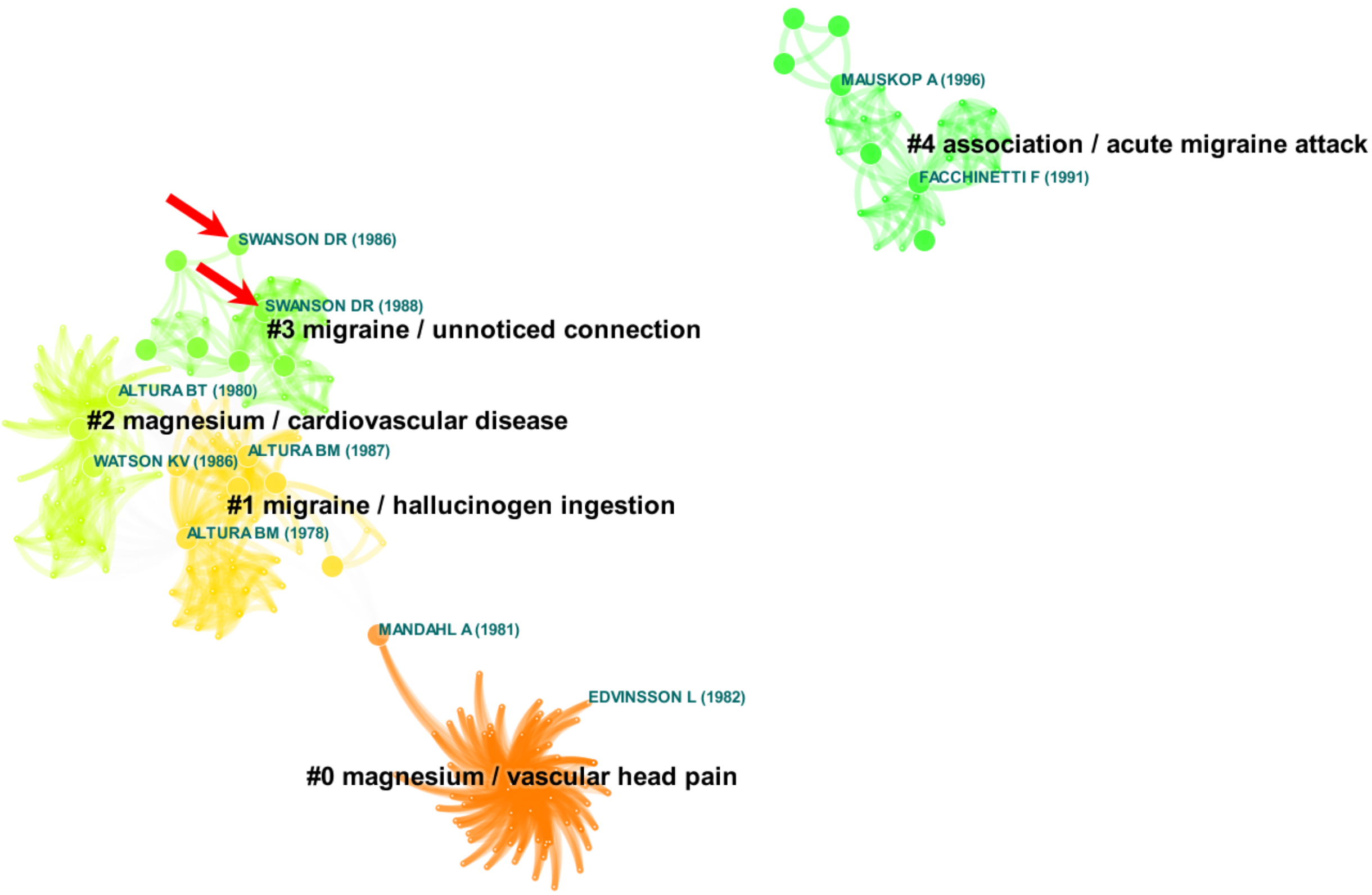

**Figure 14. The top-5 second-level clusters of the top-level cluster #15.**

*The Original Theme: Fish Oil and Raynaud's Syndrome*

Both fish oil and Raynaud's syndrome appeared in the title of Swanson's groundbreaking article (16). As shown in Figure 8, cluster #10 is on fish oil and cluster #45 is on Raynaud's syndrome.

In Figure 17, the fish oil cluster (#10) is further divided into second-level clusters. The visualized network reveals tightly coupled sub-clusters such as $\#10^0$ fish oil, $\#10^2$ blood viscosity, and $\#10^4$ chemical structure. The concept tree is represented by a single branch of fish oil and it is further split into paths such as eicosapentaenoic acid – ischaemic heart disease, and dietary supplementation – blood viscosity.

As shown in Figure 8, clusters #10 and #45 are closely connected. The area was shared by three of the five individual networks, namely $S_3$, $S_5$, and $N_B$. Figure 18 shows sub-clusters of the Raynaud's syndrome cluster (#45). The two most cited members of cluster $\#45^0$ Raynaud's syndrome were published in 1980 and 1981, respectively. The two most cited members of cluster $\#45^1$ Raynaud's phenomenon were both published in 1979. These references were published about 5 years or more before Swanson's 1986 study. These examples illustrate that a forward citation expansion may bring in older publications as well as newer ones because a new publication may cite publications older than what the original set of articles cite.

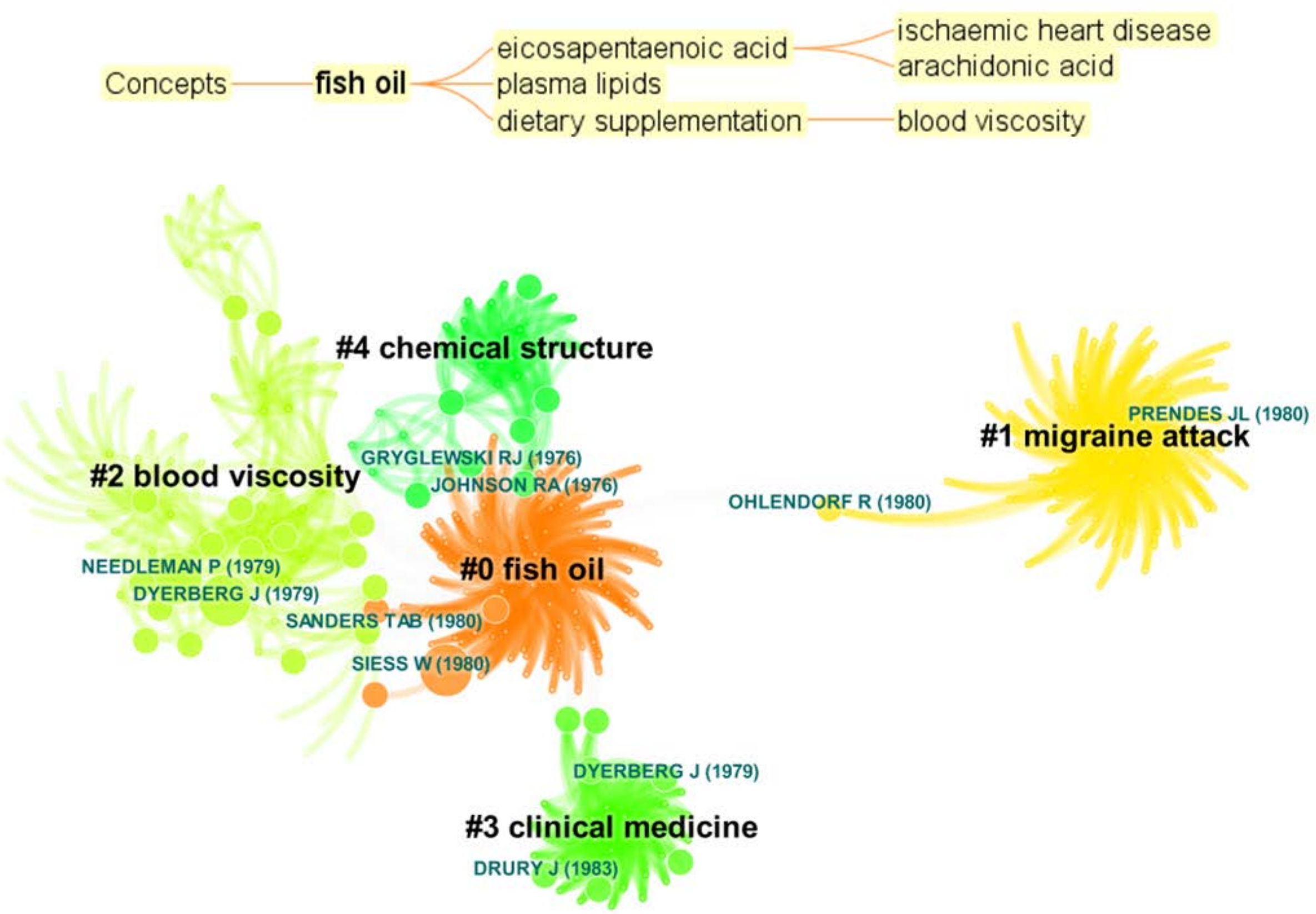

**Figure 15. Top-5 second-level clusters of top-level cluster #10 – the fish oil cluster.**

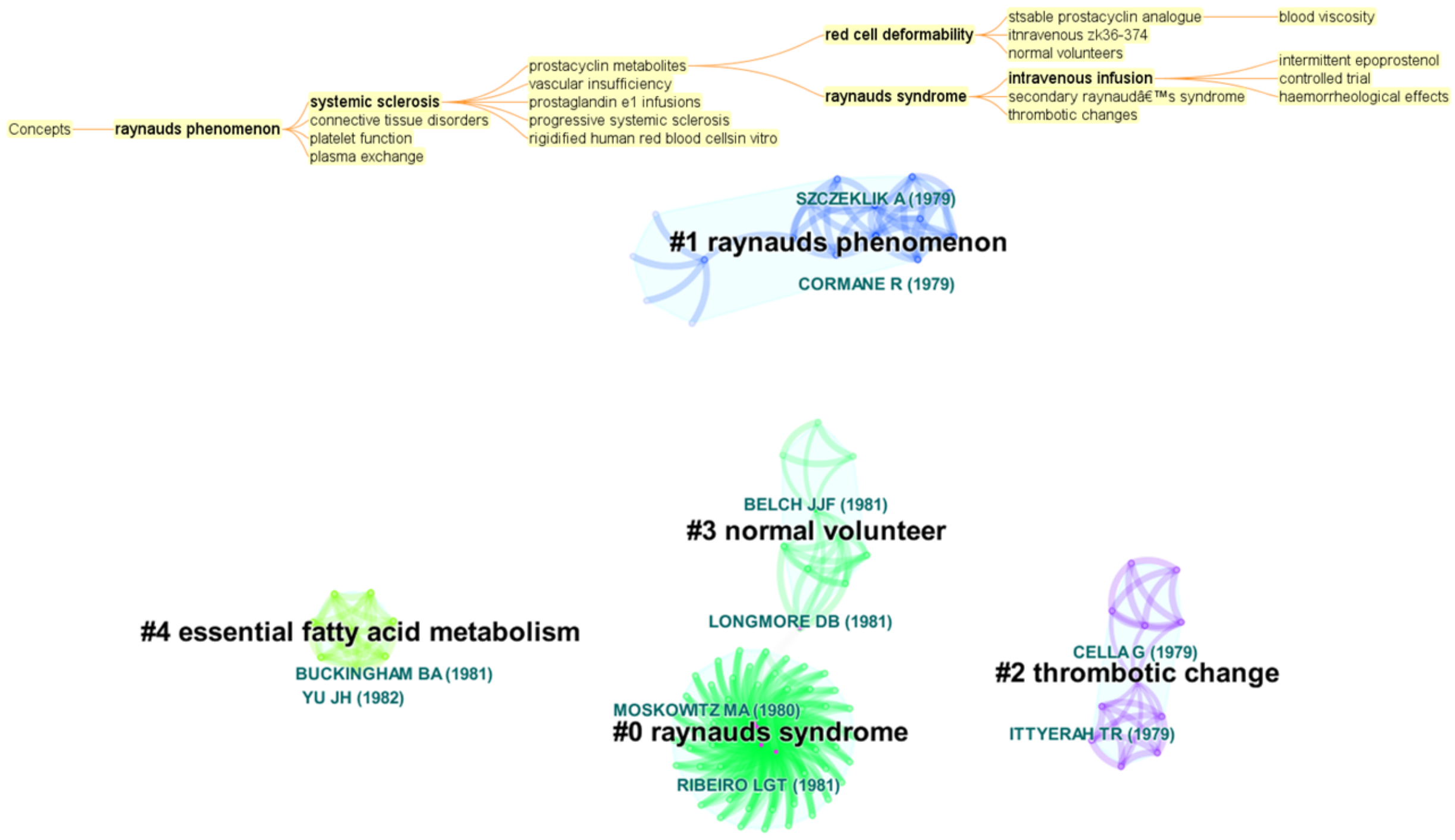

**Figure 16. Top-5 second-level clusters of top-level cluster #45 Raynaud's phenomenon.**

*What Could be Missed with a Full Text Search Alone? Deep Learning*

Recall in Figure 12, $S_5$ is the only network overlay that fully covered #32 deep learning. This example may show us what we would miss if we were to use the full text search alone. Figure 17 shows the inner structure of the cluster. The largest second-level cluster is $\#32^0$ deep learning, followed by $\#32^1$ convolutional neural network, commonly known as CNN. The concept tree consists of two branches: deep learning approaches and drug design. Several medical related concepts appear further down in the deep learning approaches branch, including medical imaging and electronic medical records, although other concepts are related to deep learning in general. If we were to rely on the full text search only, we might miss this interaction between deep learning and drug design.

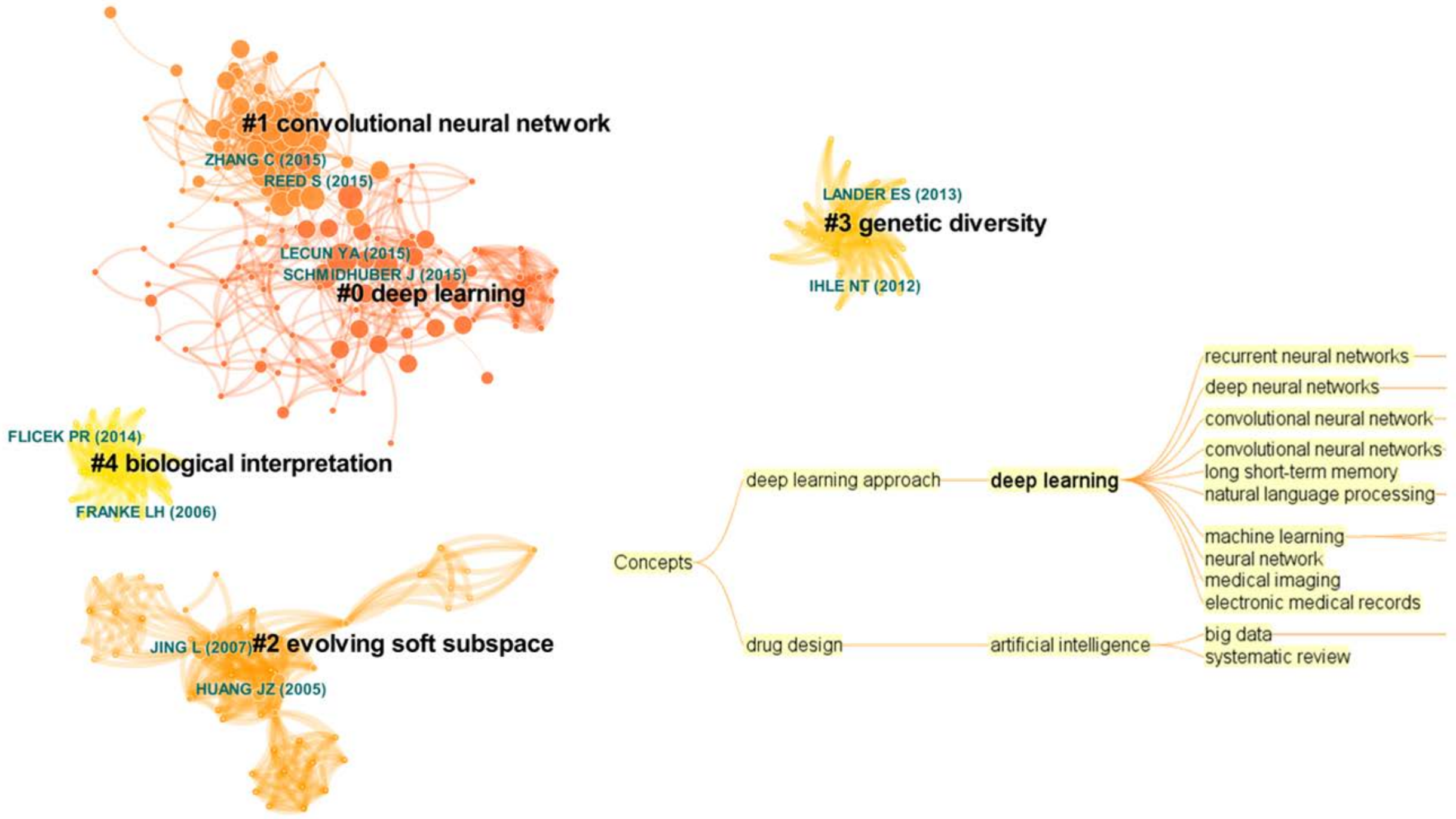

**Figure 17. Top-5 second-level clusters of top-level cluster #32 deep learning.**

# Discussions and Conclusions

We have proposed a cascading expansion method to improve the quality of systematic scientometric reviews. In particular, we have demonstrated the flexibility and extensibility of the approach that enables end users start their search process with either a pioneering article of the field or a recent review of the field. The comparisons of network visualization overlays of five datasets have revealed what a commonly used full text search strategy could have missed. Such omissions are likely to be recently emerged topics and missing them in a systematic review may undermine its overall quality. A strategy that combines query-based search and cascading citation expansion is likely to provide a more balanced coverage of a research domain and to reduce the risk of overly relying on topics explicitly specified in the initial queries. A practical implication on finding a representative body of the literature is its potential to uncover emerging topics that are currently connected to the main body of the literature through a chain of weak links. We recommend researchers to consider this strategy in situations when they only have a small number of relevant articles to begin with. As our study demonstrated, a wide variety of articles can serve as a starting point of an expansion process and multiple processes can be utilized and the combination of their results is likely to provide a comprehensive coverage of the underlying thematic landscape of a research field or a discipline.

The present study has some limitations and it raises new questions that need to be addressed in future studies. Our approach implies an assumption that the structure of scientific knowledge can be essentially captured through semantically similar text and/or explicit citation links. Is this assumption valid at the disciplinary level? To what extent does the choice of the seed articles for the expansion process matter? Does the choice of seed articles influence the stability of the expansion process? How many generations of expansion would be optimal?

We have made a number of observations and recommendations that are potentially valuable for adapting this type of search strategies to develop a systematic review of a body of scientific literature of interest.

- Using a combination of multiple cascading citation expansions with different seed articles is recommended to obtain a more balanced representation of a field than using a full text search alone.
- Multi-generation citation expansions provide a systematic approach to reduce the risk of missing topics that we may not be familiar with or not aware of altogether.
- Areas where multiple networks overlap indicate core topics of the field of research, namely literature-based discovery, text mining, big data, genomic era, and domain knowledge.
- Multiple levels of network analysis and cluster analysis are recommended, in consistent with information strategies such as scatter/gatherer. One may find additional insights by decomposing a top-level cluster into second-level clusters.
- Choosing the starting point and an end point of a cascading expansion process may lead to different results, suggesting the complexity of the networks and threshold selections may play important roles in reproducing the results in similar studies.
- Modularity and cluster silhouette measures can help us to assess the quality of an expansion process.

Comparing multiple networks in the same context allows us to identify the topic areas that are particularly well represented in some of the datasets but not in other ones. Such an understanding of the landscape of a field provides additional insights into the structure and the long-term development of the field.

As a methodology for generating systematic scientometric reviews of a knowledge domain, it bridges the formally mutual exclusive globalism and localism by providing a scalable transition mechanism between them. The most practical contribution of our work is the development and dissemination of a tool that is readily accessible by end users.

## Notes

CiteSpace is available at https://sourceforge.net/projects/citespace/.

The five datasets are available on Google Drive at https://tinyurl.com/yyj9hhph.

## Acknowledgements

The work is supported by the SciSIP Program of the National Science Foundation (Award #1633286). CM acknowledges the support of Microsoft Azure Sponsorship. Data sourced from Dimensions, an inter-linked research information system provided by Digital Science (https://www.dimensions.ai). This work is also supported by the Ministry of Education of the Republic of Korea and the National Research Foundation of Korea (NRF-2018S1A3A2075114). This research is also partially supported by the Yonsei University Research Fund of 2019-22-0066.

## References

1. Price DD. Networks of scientific papers. Science. 1965;149:510-5.
2. Cobo MJ, López-Herrera AG, Herrera-Viedma E, Herrera F. Science mapping software tools: Review, analysis, and cooperative study among tools. J Am Soc Inf Sci Technol. 2011;62(7):1382-402.
3. Chen C. CiteSpace II: Detecting and visualizing emerging trends and transient patterns in scientific literature. J Am Soc Inf Sci Technol. 2006;57(3):359-77.

4. Chen C. Science Mapping: A Systematic Review of the Literature. Journal of Data and Information Science. 2017;2(2):1-40.
5. van Eck NJ, Waltman L. Software survey: VOSviewer, a computer program for bibliometric mapping. Scientometrics. 2010;84(2):523-38.
6. Chen C, Hu Z, Liu S, Tseng H. Emerging trends in regenerative medicine: A scientometric analysis in CiteSpace. Expert Opinions on Biological Therapy. 2012;12(5):593-608.
7. Shen S, Cheng C, Yang J, Yang S. Visualized analysis of developing trends and hot topics in natural disaster research. PLoS One. 2018;13(1):e0191250.
8. Zhang C, Xu T, Feng H, Chen S. Greenhouse Gas Emissions from Landfills: A Review and Bibliometric Analysis. Sustainability. 2019;11(8):2282.
9. Li M, Porter AL, Suominen A. Insights into relationships between disruptive technology/innovation and emerging technology: A bibliometric perspective. Technological Forecasting and Social Change. 2018;129:285-96.
10. Kullenberg C, Kasperowski D. What Is Citizen Science? – A Scientometric Meta-Analysis. PLoS One. 2016;11(1):e0147152.
11. Haunschild R, Bornmann L, Marx W. Climate Change Research in View of Bibliometrics. PLoS One. 2016;11(7):e0160393.
12. Klavans R, Boyack KW. Using Global Mapping to Create More Accurate Document-Level Maps of Research Fields. J Am Soc Inf Sci Technol. 2011;62(1):1-18.
13. Leydesdorff L, Rafols I. A Global Map of Science Based on the ISI Subject Categories. J AM SOC INF SCI TEC. 2009;60(2):348-62.
14. Borner K, Klavans R, Patek M, Zoss AM, Biberstine JR, Light RP, et al. Design and Update of a Classification System: The UCSD Map of Science. PLoS One. 2012;7(7):10.
15. Garfield E. From the science of science to Scientometrics visualizing the history of science with HistCite software. Journal of Informetrics. 2009;3(3):173-9.
16. Swanson DR. Fish oil, Raynaud's syndrome, and undiscovered public knowledge. Perspectives in Biology and Medicine. 1986;30(1):7-18.
17. Swanson DR. Undiscovered public knowledge. Library Quarterly. 1986;56(2):103-18.
18. Swanson DR. 2 MEDICAL LITERATURES THAT ARE LOGICALLY BUT NOT BIBLIOGRAPHICALLY CONNECTED. J Am Soc Inf Sci. 1987;38(4):228-33.
19. Swanson DR. MIGRAINE AND MAGNESIUM - 11 NEGLECTED CONNECTIONS. Perspectives in Biology and Medicine. 1988;31(4):526-57.
20. Smalheiser NR, Swanson DR. Using ARROWSMITH: a computer-assisted approach to formulating and assessing scientific hypotheses. Comput Meth Programs Biomed. 1998;57(3):149-53.
21. Swanson DR, Smalheiser NR. An interactive system for finding complementary literatures: A stimulus to scientific discovery. Artificial Intelligence. 1997;91(2):183-203.
22. Smalheiser NR, Swanson DR. Linking estrogen to Alzheimer's disease: An informatics approach. Neurology. 1996;47(3):809-10.
23. Smalheiser N. Rediscovering Don Swanson: the past, present and future of literature-based discovery. Journal of Data and Information Science. 2017;2(4):43-64.
24. Weeber M, Klein H, de Jong-van den Berg LTW, Vos R. Using concepts in literature-based discovery: Simulating Swanson's Raynaud-fish oil and migraine-magnesium discoveries. J Am Soc Inf Sci Technol. 2001;52(7):548-57.
25. Gordon MD, Lindsay RK. Toward discovery support systems: A replication, re-examination, and extension of Swanson's work on literature-based discovery of a connection between Raynaud's and fish oil. J Am Soc Inf Sci. 1996;47(2):116-28.
26. Kim YH, Song M. A context-based ABC model for literature-based discovery. PLoS One. 2019;14(4).
27. Kostoff RN. Literature-related discovery (LRD): Introduction and background. Technological Forecasting and Social Change. 2008;75(2):165-85.

28. Shneider AM. Four stages of a scientific discipline: four types of scientists. Trends in biochemical sciences. 2009;34(5):217-23.
29. Chen C, Song M. Representing Scientific Knowledge: The Role of Uncertainty: Springer; 2017.
30. Chen C. Predictive effects of structural variation on citation counts. J Am Soc Inf Sci Technol. 2012;63(3):431-49.
31. Klavans R, Boyack KW. Toward a Consensus Map of Science. J Am Soc Inf Sci Technol. 2009;60(3):455-76.
32. Chen C, Leydesdorff L. Patterns of connections and movements in dual-map overlays: A new method of publication portfolio analysis. J Am Soc Inf Sci Technol. 2014;62(2):334-51.
33. Chen C. Searching for intellectual turning points: Progressive knowledge domain visualization. Proc Natl Acad Sci U S A. 2004;101:5303-10.
34. Börner K, Chen C, Boyack KW. Visualizing knowledge domains. Annual Review of Information Science and Technology. 2003;37(1):179-255.
35. Pirolli P, Schank P, Hearst M, Diehl C, editors. Scatter/Gather browsing communicates the topic structure of a very large text collection. the Conference on Human Factors in Computing Systems (CHI '96); 1996 April 1996; Vancouver, BC: ACM Press.
36. Miller GA. WORDNET - A LEXICAL DATABASE FOR ENGLISH. Communications of the Acm. 1995;38(11):39-41.
37. Aronson AR, Lang FM. An overview of MetaMap: historical perspective and recent advances. Journal of the American Medical Informatics Association. 2010;17(3):229-36.
38. Bodenreider O. The Unified Medical Language System (UMLS): integrating biomedical terminology. Nucleic Acids Research. 2004;32:D267-D70.
39. Deerwester S, Dumais ST, Landauer TK, Furnas GW, Harshman RA. Indexing by Latent Semantic Analysis. J Am Soc Inf Sci. 1990;41(6):391-407.
40. Mikolov T, Sutskever I, Chen K, Corrado GS, Dean J. Distributed representations of words and phrases and their compositionality. Proceedings of the 26th International Conference on Neural Information Processing Systems (NIPS'13); December 05 - 10, 2013; Lake Tahoe, Nevada2013. p. 3111-9.
41. Bush V. As we may think. The Atlantic Monthly. 1945;176(1):101-8.
42. Garfield E. Citation indexing for studying science. Nature. 1970;227:669-71.
43. Merton RK. Priorities in scientific discoveries. American Sociological Review. 1957;22:635-59.
44. Chen C. Cascading citation expansion. Journal of Information Science Theory and Practice. 2018;6(2):6-23.
45. Chen C, Ibekwe-SanJuan F, Hou JH. The Structure and Dynamics of Cocitation Clusters: A Multiple-Perspective Cocitation Analysis. J Am Soc Inf Sci Technol. 2010;61(7):1386-409.
46. Swanson DR. Migraine and magnesium: Eleven neglected connections. Perspectives in Biology and Medicine. 1988;31(4):526-57.